\documentclass[12pt]{article}

\usepackage[french,english]{babel} 
\usepackage[utf8]{inputenc} 
\usepackage[T1]{fontenc} 
\usepackage{lmodern}
 
\usepackage{float}  

\usepackage{longtable}
\usepackage{dcolumn}

\usepackage{framed}

\usepackage{amsthm,amssymb,mathrsfs,bm}
\usepackage[tbtags]{amsmath}
\usepackage{subeqnarray}
\usepackage{eurosym}

\usepackage{relsize,exscale}
\usepackage{tabularx}
\usepackage{multirow}
\usepackage{booktabs}
\usepackage{graphicx}
\usepackage{setspace}
\usepackage{lscape}
\usepackage{vmargin}
\usepackage{color}

\usepackage{mdframed} 

\definecolor{dark-blue}{RGB}{21,85,212}
\definecolor{light-blue}{RGB}{51,153,255}
\definecolor{dark-green}{RGB}{50,150,86}
\definecolor{dandelion}{RGB}{212,174,21}
\definecolor{bloud}{RGB}{212,21,21}
\definecolor{NewPurple}{RGB}{127,0,255}

\usepackage{url}
\usepackage{enumerate}
\usepackage{fancyhdr}

\usepackage{natbib}

\renewenvironment{thebibliography}[1]{%
\begin{oldthebibliography}{#1}
    \setlength{\parskip}{0.00mm}
    \setlength{\itemsep}{0.00mm}}
{\end{oldthebibliography}}

\usepackage{hyperref}
\hypersetup{
    colorlinks=true,
    linkcolor=blue,
    filecolor=magenta,      
    urlcolor=blue,
    citecolor=blue,
    pdftitle={Overleaf Example},
    pdfpagemode=FullScreen,
    }

\usepackage[hang,splitrule]{footmisc} 


\usepackage[sf,bf,tiny,center]{titlesec} 
\titlelabel{\thetitle.\enspace}
\usepackage{titletoc}

\usepackage{nicefrac}

\renewcommand{\thefootnote}{\fnsymbol{footnote}}
\setmarginsrb{50pt}{50pt}{50pt}{70pt}{20pt}{20pt}{20pt}{30pt}

\newcommand{\x}{\bm x}
\newcommand{\y}{\bm y}

\newcommand{\xt}{\bm t}

\newcommand{\real}{\mathbb R}

\newcommand{\income}{\mathscr D}
\newcommand{\ensn}{\mathscr{D}^{n}}

\newcommand{\ensU}{\mathscr U}

\newcommand{\ensF}{\mathscr F}

\DeclareMathOperator*{\defn}{=}

\newtheorem{proposition}{Proposition}

\newtheorem{definition}{Definition}

\title{Does the Gini index represent\\ people's views on inequality?\,\footnotemark[1]}

\author{Gaëlle Aymeric,\footnotemark[2]\ \ and Brice Magdalou\,\footnotemark[3]}

\begin{document}
\onehalfspacing
\maketitle

\footnotetext[1]{This paper is part of the research projects \textit{RediPref} (Contract ANR-15-CE26-0004) and \textit{MaDimIn} (Contract ANR-24-CE26-3823-02), from which financial support is acknowledged. We are grateful to Elena B\'arcena-Mart\'in, Thierry Blayac, Frank Cowell, Patrick Moyes, Vito Peragine, Daniel Sant\'in, Rafael Salas, Erik Schokkaert, Benoit Tarroux and Alain Trannoy for interesting discussions related to the experimental approach used in this paper, and the results. Most of this paper was written when Brice Magdalou was on a temporary CNRS research assignment at the Aix-Marseille School of Economics.}
\footnotetext[2]{CEE-M, Univ. Montpellier, CNRS, INRAE, Institut Agro, Montpellier, France. \\ E-mails: \texttt{gaelle.aymeric@umontpellier.fr}.}
\footnotetext[3]{CNRS, AMSE, Marseille, France. \\ E-mail: \texttt{brice.magdalou@umontpellier.fr}.}


\renewcommand{\thefootnote}{\arabic{footnote}}

\pagestyle{fancy}
\fancyhf{}
\chead{\textsc{Does the Gini index represent people's views on inequality?}}
\renewcommand{\headrulewidth}{0pt}
\cfoot{\thepage} 

\begin{abstract}

This paper presents findings from a web-experiment on a representative sample of the French population. It examines the acceptability of the Pigou-Dalton principle of transfers, which posits that transferring income from an individual to a relatively poorer one, reduces overall inequality. While up to 60\% of respondents reject standard transfers, the three alternative transfers we test receive more approval, especially those promoting solidarity among lower-income recipients. The study then models respondents' preferences with two types of social welfare functions, utilitarian and Extended Gini. The Extended Gini model aligns better with individual preferences. Nevertheless, Extended Gini-type social welfare functions that adhere to the principle of transfers (including the one underlying the Gini index) poorly capture preferences of each individual. However, quite surprisingly, the preferences of the median individual align almost perfectly with the Gini-based function, using either parametric or non-parametric estimates.

\vspace{0.4cm}
\end{abstract}

\newpage

\section{Introduction}
\label{intro}

Since the seminal work of \cite{K69} and \cite{At70}, the \textit{Pigou-Dalton principle of transfers} is the cornerstone of the theory of income inequality measurement. According to this normative principle, a mean-preserving transfer of income from one individual to another who is relatively poorer, without reversing the initial positions on the income scale, always reduces overall income inequality. Almost all the income inequality indices used today, both in academic research and by official statistical institutes, conform to this principle. The Gini index, which is by far the most widely used, is a prime example. 

In this paper, we look at whether this approach is compatible with people's views on inequality. We report the results of a web-experiment conducted on a representative sample of the French population, with 1,028 participants. Subjects were asked to compare, in terms of inequality, pairs of income distributions for an hypothetical society, where all individuals are clones. We test the acceptability of the principle of transfers, but also of three alternative principles that impose constraints on recipients and donors in transfers. Then, assuming that individual preferences of participants can be represented by a social welfare function we estimate, parametrically and non-parametrically, two functions commonly used in the literature: \textit{utilitarianism} and \textit{extended Gini}. Since participants' preferences are heterogeneous, we focus on the preferences of the \textit{median individual}. We then compare our estimates with the social welfare functions used in the empirical literature, in particular the one underlying the Gini index. To our knowledge, this is the first study of its kind involving a representative sample of a country's population, and testing alternatives to the principle of transfers. It is also the first experimental study which investigates the extended Gini model, making it possible to test the relevance of Gini as an index representing people's views on inequality.

\begin{figure}[!htp] \caption{\label{Trans-Illustr} \textit{Impact of a progressive transfer on the global distribution}} 
\small
\begin{tabular}{cc}
\includegraphics[width=8.5cm]{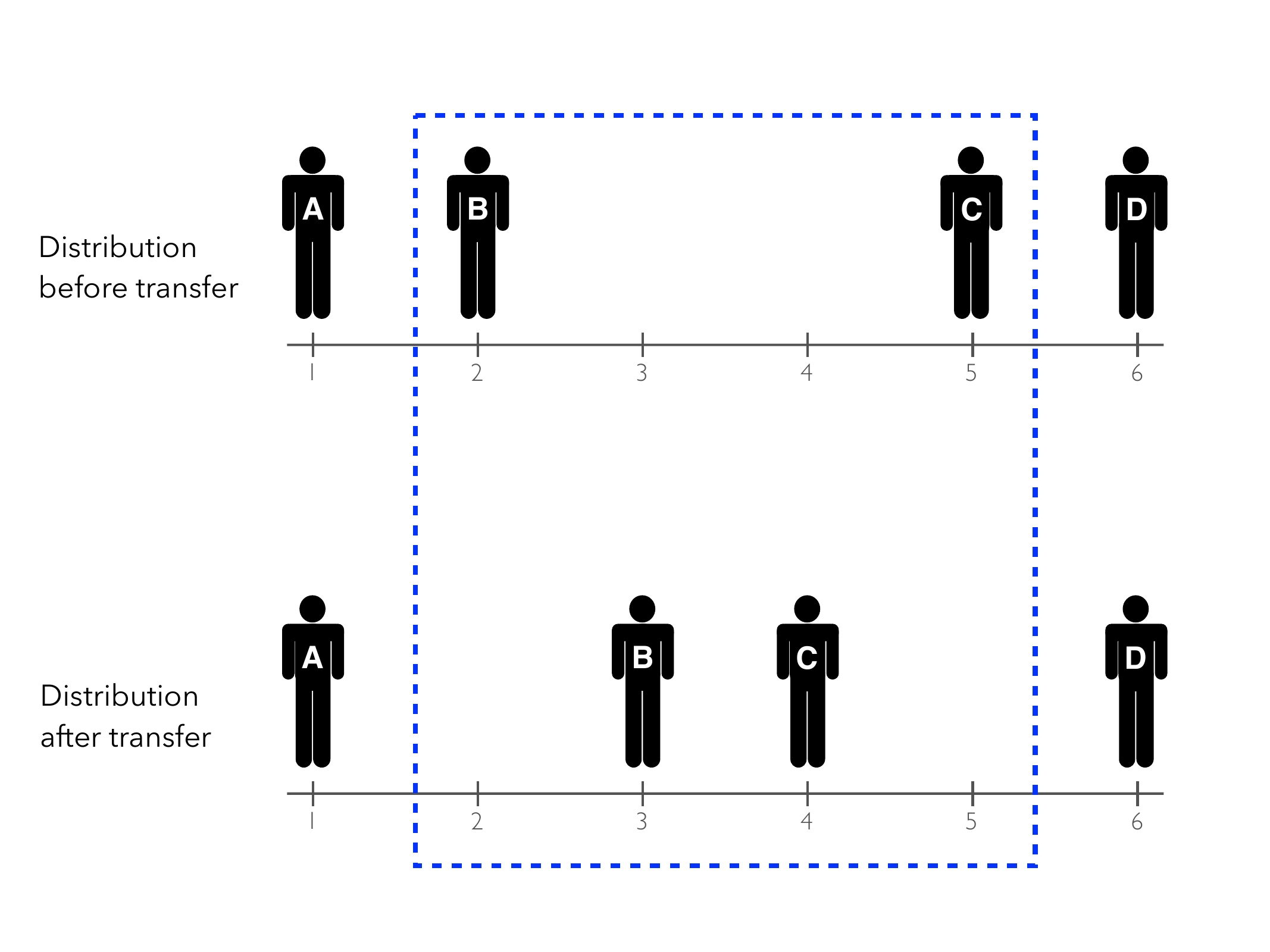} & %
\includegraphics[width=8.5cm]{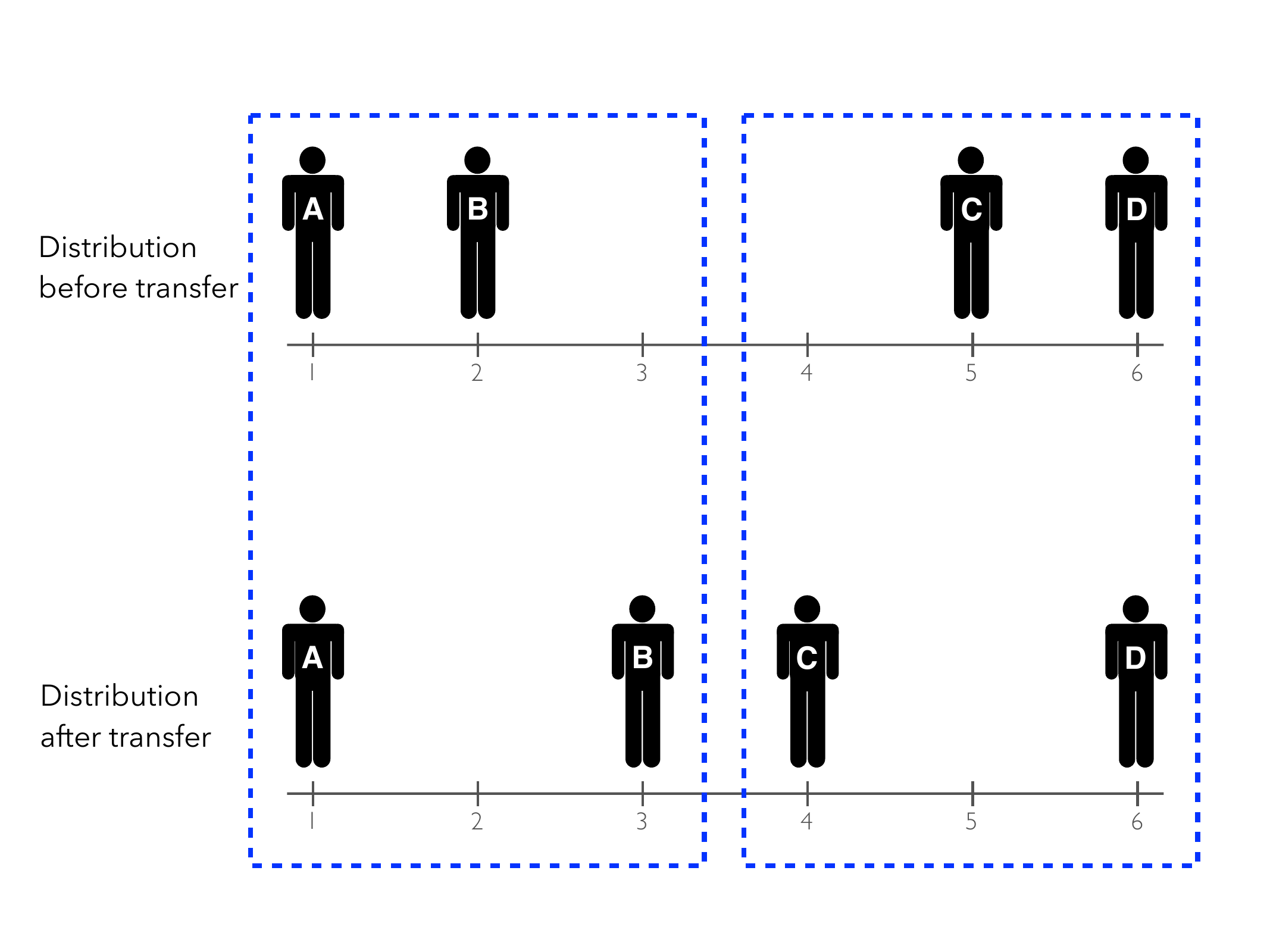} \\ 
\end{tabular}

\end{figure}

Why should the principle of transfers pose a problem? Obviously, income inequality is reduced between the two people involved in a progressive transfer. However, the effect of such a transfer on the overall inequality in the distribution may be open to discussion. By way of illustration, let's consider a society consisting of 4 individuals who are perfectly identical, apart from their income. Let's call these individuals A, B, C and D and say that they have, respectively, 1, 2, 5 and 6 income units. Suppose a transfer of 1 unit of income is made from individual C to B. The left-hand side of Figure~\ref{Trans-Illustr} illustrates the fact that these two individuals get closer on the income scale. 
The (relative) situation of the individuals not concerned by the transfer is highlighted in the right-hand side. After the transfer, the poorest individual A is left behind. Even if she can be considered a priority target for redistribution, her situation remains unchanged, whereas it is improved for individual B, who was already richer than her before the transfer. On the other side of the distribution, individual C contributes to the transfer, while the richest individual D does not. For these reasons, some people may consider that a distribution obtained by means of certain progressive transfers is not necessarily more equal than the distribution before transfer.

The previous example illustrates a deep-seated problem with the principle of transfers: It associates the notion of `reduction of a partial statistic distance' to the notion of `reduction in global inequality'. As inequality is a normative concept, the equivalence between these two notions is not immediate. \cite{AC92} were the first to observe, through a questionnaire experiment with students, the low approval of certain progressive transfers, particularly those not involving individuals at the extremes of the distribution. Despite the fact that this finding has been confirmed by numerous subsequent experiments \cite[][to name a few]{AC02, GN03, ACG12}, few studies have sought to find an alternative to this principle that is more in line with people's preferences. 

This low approval rate of the principle of transfers may be explained by the way people assess their own position in the income distribution (even if they are not involved in the distributions they have to compare). Although individuals attach importance to the absolute level of income, it is now generally accepted that they attach as much, if not more, importance to the \textit{relative position} on the income scale \cite[][]{CFS08}. They are also sensitive to changes in their own situation, relative to individuals in their \textit{reference group}, which can be defined on the basis of social, professional or neighborhood considerations. In various branches of economic literature (macroeconomics, finance, labour economics, etc.), this robust fact has been referred to the \textit{keeping up with the Joneses} effect \cite[]{Ab90}. Comparable considerations are also found, for example, in sociology through the notion of \textit{deprivation} \cite[][]{Ru66}, or in philosophy with the notion of \textit{complaint} \cite[][]{Te86, Te93}. In the example given in Figure~\ref{Trans-Illustr}, we can see that the initial distribution is in some sense polarised, with a homogeneous group of two individuals at the bottom of the distribution, and the same at the top. After the transfer, individual B's income gain is associated with a deterioration in the relative situation of individual A in the bottom group, and individual C in the top group. These cumulative changes may therefore not be seen as reducing overall inequality. 

\cite{CM06} proposed three possible alternatives to the principle of transfers. This principle (PT hereafter) imposes no restrictions on the income transfers, other than the fact that the transfer must be mean-preserving, and that the donor must be and remain richer than the recipient. The alternative principles impose a form of solidarity (or uniformity) at the bottom and/or the top of the distribution. At the bottom of the distribution, if an individual receives a certain amount of income, the same amount at least must be received by those poorer than him. This is called a \textit{uniform-on-the-left} transfer (UL). At the top of the distribution, if an individual is a donor, all those richer than him must contribute, at least as much. This is called a \textit{uniform-on-the-right} transfer (UR). A transfer that combines the two restrictions is called a \textit{uniform-on-the-right-and-left} transfer (URL).
These authors then identified the preorder relations (comparable to the Lorenz criterion for the principle of transfers) and the restrictions to be placed on the extended Gini class of social welfare functions, to make inequality assessment compatible with these new principles.

Our results can be summarised as follows. If we focus on the numerical questions, where participants have to compare pairs of distributions, the Pigou-Dalton principle of transfers is validated in only 38\% of cases (Table~\ref{Resultats-globaux}). This result, which is very low, is comparable to that observed in the literature. Uniform transfers receive significantly more support, up to 62\% for URL transfers. If uniformity is imposed on only one side of the distribution, UL transfers (solidarity among the poor) seem to be preferred. We then focus on the sub-sample of participants who correctly answered the 4 test questions we set up (to ensure they understood the experiment). The trend observed in the full sample is confirmed (Table~\ref{Resultats-globaux-sans-erreur}), but with an increase in acceptance rate of all transfers (from 52\% for the principle of transfers to 67\% for URL). This trend remains robust, whatever the initial distribution considered before transfer, and whatever the socio-economic characteristics of the participants (gender, level of education, political views, employment status). We find similar results in questions where transfers (with or without solidarity) are described in text form (Table~\ref{Resultats-globaux-text}), but with smaller differences in acceptance rates (UR and UL are not significantly different here).

The econometric estimation of social welfare functions for each participant also provides interesting results. We first compare the parametric estimations of the utilitarian function underlying the \textit{Atkinson-Kolm-Sen inequality indices} \cite[][]{K69, At70, Se73}, to the weighting function underlying the \textit{extended Gini indices} \cite[][]{Y87, DW80}. The two functions are one-parameter power functions. Consistency with the principle of transfers is captured in the two models, respectively, by the concavity of the utility function and the convexity of the weighting function. This is what we observe for over 80\% of participants in the restricted sample (Table~\ref{Resultats-shape}). The median value of the utility function parameter varies according to the estimation method (Table~\ref{Resultats-Utility}), while that of the weighting function always remains very close to that characterising the Gini index (Table~\ref{Resultats-extended-Gini} and Figure~\ref{Param-estim}). If we compare the two approaches using the \textit{Akaike Information Criterion}, we see that the extended Gini model fits the data better for 70\% of participants (Table~\ref{Resultats-AIC}). Hence, the extended Gini model seems to have an empirical advantage here. \cite{CM06} have also shown that this model is sufficiently flexible to distinguish between the different transfer principles (URL, UL, UR, PT), which is not possible for the utilitarian model. In the latter case, consistency with all the principles always comes down to the concavity of the utility function. We provide, in a second step, a non-parametric (point-by-point) estimation of the weighting function for each participant. Our descriptive results are confirmed: the classes of functions compatible with uniform transfers represent a larger proportion of participants, with a clear hierarchy: URL > UL > UR > PT, ranging from [76\%-77\%] of the participants for URL to [10\%-36\%] for PT, depending on the estimation methods (Table~\ref{Resultats-f-classes}). Nevertheless, for the median individual, we obtain a result similar to the parametric estimate: the weighting function characterising the Gini index approximates fairly closely that of the median individual (Figure~\ref{Non-param-estim}).

The rest of the paper is organised as follows. In Section~\ref{sec-theory}, we outline the normative theory of inequality measurement, based on the  Pigou-Dalton principle of transfers. We set out the notation, the main income inequality measurement tools and the alternatives to this principle. We also discuss some experimental results which, although they do not address precisely the same issues as we do, seem to confirm the empirical relevance of the new principles. We present our experimental design in Section~\ref{sec-design}. We detail the successive stages of the web-experiment, and we show how the pairs of income distributions under comparison are constructed. The sample of participants is described in Section~\ref{sec-sample}. Section~\ref{sec-acceptance} discusses the acceptance rates for the different transfer principles. We focus first on the numerical questions and we look at whether the initial distributions, or the socio-economic characteristics of the participants, have an influence on these acceptance rates. We then present the acceptance rates in the text-based questions. The econometric estimations of the social welfare function for each participant is detailed in Section~\ref{sec-sw}, by distinguishing the parametric and non-parametric approaches. Finally, Section~\ref{sec-discussion} concludes.  

\section{The normative approach to measuring income inequality}
\label{sec-theory}

\subsection{The framework}
\label{sec-framework}

\noindent \textsc{Notation}. We consider a population consisting of $n \geq 2$ individuals, identical in every respects other than their income. Income of individual $i$ is denoted by $x_i \in \income$, where $\income$ is an interval of the non-negative real numbers, and an income distribution is a list $\x = (x_1,x_2,\dots,x_n)$. We restrict attention to non-decreasingly ordered distributions, and the set of these distributions is denoted~$\income^n$.\footnote{This framework is deliberately simplified, and it is defined without lost of generality. The inequality measures we present in this paper are all consistent with the symmetry axiom (invariance with respect to a permutation of the distribution) and Dalton's principle of populations (invariance to an identical replication of the population).} The mean income of distribution~$\x$ is indicated by $\mu(\x) \defn \sum_{i=1}^{n} x_i / n$.

\noindent \textsc{Inequality reduction}. 
In the inequality measurement literature, it is usually assumed that a mean-preserving transfer of income from one individual to another who is relatively poorer, without reversing the initial positions on the income scale, always reduces inequality.  Such a transfer is known as a Pigou-Dalton progressive transfer, and is formally defined as follows:
\begin{definition}[Pigou-Dalton Transfer] \label{ULPT} Given two income distributions $\x, \y \in \ensn$, we say that $\x$ is obtained from $\y$ by means of a Pigou-Dalton progressive transfer, if there exists $\delta > 0$ and two individuals $1 \leq h<k \leq n$ such that $\x = \y + \xt$ and:
%
%
\begin{center}
\textnormal{
\small
\begin{tabular}{r*{11}{c}}
        \toprule
\textsc{Individual} = & $1$ & \dots & $h-1$  & $h$       & $h+1$ & \dots & $k-1$ & $k$         & $k+1$ & \dots & $n$  \\
$\xt =$              & ($0$ & \dots & $0$    & $\delta$ & 0         & \dots & 0        & $-\delta$ & $0$     & \dots & $0$) \\
\bottomrule
\end{tabular}
}
\end{center}
\end{definition}

\vspace{0.2cm}
\noindent Acceptance of the inequality-reducing nature of such a transfer is known as the Pigou-Dalton principle of transfers.

\noindent \textsc{Social welfare functions}. According to the so-called ethical approach to measuring inequality \cite[][]{BBD99}, it is used to assess the inequality of a distribution $\x \in \income^n$ on the basis of a social welfare function $W:\income^n \longrightarrow \real$.\footnote{A relative inequality index can be written as $I(\x) = 1 - \Xi(\x) / \mu(\x)$, where the \textit{equally distributed equivalent income} $\Xi(\x)$ is the income which, if received by each individual, gives rise to a distribution socially indifferent to~$\x$. $\Xi(\x)$ is implicitely defined by $W (\x) = W (\Xi(\x), \dots, \Xi(\x))$.} Traditionally two subclasses of the \textit{rank-dependent expected utility model} popularized by \cite{Qu93} are considered in the literature. Firstly, there is the \textit{utilitarian approach}, which assumes that social welfare is the average of the utilities obtained by individuals, denoted by:
\begin{equation} \label{scw-utili}
W_u(\x) = \frac{1}{n} \sum_{i=1}^{n} u(x_i)\,,\quad \forall \x \in \income^n\,,
\end{equation}
where $u \in \ensU \defn \{ u:\income \longrightarrow \real \mid
u\ \textnormal{continuous and non-decreasing} \}$ is the utility function (defined up to an increasing affine transformation). Secondly, there is the \textit{extended Gini approach}, with the social welfare function:
\begin{equation} \label{scw-yaari1}
W_f(\x) = \sum_{i=1}^{n} \bigg[ f \bigg(\frac{n-i+1}{n} \bigg)-f
\bigg(\frac{n-i}{n} \bigg) \bigg] x_i\,,\quad \forall \x \in \income^n\,,
\end{equation}
where $f \in \ensF \defn \{ f:[0,1] \longrightarrow [0,1] \mid f\ \textnormal{continuous, non-decreasing}\,,\ f(0)=0\,, f(1)=1 \}$ is the weighting function.

\noindent \textsc{Equivalence result}. The main result in the literature on income inequality measurement is the Hardy-Littlewood-Polya theorem \cite[HLP hereafter, see][]{Ma21}, which establishes the equivalence between several statements. To be precise, each statement defines a preorder relation (incomplete ranking) which describes a situation where a distribution $\x$ provides more social welfare than a distribution $\y$. Because the averages of the two distributions are equal, this also means that $\x$ is less unequal. Formally, we have:\footnote{In fact, the Hardy-Littlewood-Polya theorem is the equivalence between statements (a), (b1) and (c). The equivalence with statement (b2) can be derived from Theorem~2 in \cite{Y87}.}

\begin{proposition} \label{res-PD-transfers-ineq}
Let $\x, \y \in \ensn$ such that $\mu(\x)=\mu(\y)$. The following statements are equivalent: 
\begin{itemize}
\item[(a)] $\x$ is obtained from $\y$ by means of a sequence of Pigou-Dalton transfers,
\item[(b1)]  $W_u(\x)  \geq W_u(\y)$, for all concave functions $u \in \ensU$,
\item[(b2)]  $W_f(\x)  \geq W_f(\y)$, for all convex functions $f \in \ensF$,
\item[(c)] $\frac{1}{n} \sum_{i=1}^{h} x_i \geq \frac{1}{n} \sum_{i=1}^{h} y_i$, for all $h = 1,\ldots,n-1$.
\end{itemize}
\end{proposition}

\noindent The first statement describes an unambiguous reduction of inequality, in the sense of the principle of transfers view on inequality. The second statement presents the condition to be placed on the utility function to obtain an utilitarian social ranking of the distributions, consistent with the principe of transfers. The third statement is the same, but within the extended Gini framework. Finally, statement (c) corresponds to the well-known Lorenz criterion.

\subsection{Alternatives to the principle of transfers}
\label{subsec-alt-principle}

While a progressive transfer indisputably reduces inequality between the individuals concerned by the transfer, it is less clear that inequality is reduced in the entire distribution, as illustrated in Figure~\ref{Trans-Illustr}. \cite{CM06} have proposed three possible restrictions to be applied to a transfer to be considered as inequality reducing.

The first one is \textit{uniform transfers on the right and on the left} (URL). According to this alternative view, an income transfer reduces the inequality only if three conditions are satisfied. First, when an amount of income is taken to one individual, the same amount has to be taken to all the individuals richer than her. Symmetrically, when an individual receives an amount of income, the same amount has to be received by all the individuals poorer than her. Moreover, the mean income must be preserved after transfer. By definition, an URL transfer is a (sequence of) progressive transfer(s), but most of the progressive transfers are not URL transfers. 

\begin{definition}[URL Transfer] \label{URL} Given two income distributions $\x, \y \in \ensn$ such that $\mu(\x)=\mu(\y)$, we say that $\x$ is obtained from $\y$ by means of a uniform-on-the-right-and-left progressive transfer, if there exist $\delta, \epsilon > 0$ and two individuals $1 \leq h<k \leq n$ such that $\x = \y + \xt$ and:
%
%
\begin{center}
\textnormal{
\small
\begin{tabular}{r*{11}{c}}
        \toprule
\textsc{Individual} = & $1$         & \dots & $h-1$        & $h$       & $h+1$ & \dots & $k-1$ & $k$             & $k+1$            & \dots & $n$             \\
$\xt =$              & ($\delta$ & \dots & $\delta$    & $\delta$ & 0         & \dots & 0        & $-\epsilon$ & $-\epsilon$    & \dots & $-\epsilon$) \\
\bottomrule
\end{tabular}
}
\end{center}
\end{definition}

\vspace{0.2cm}
\noindent The second restriction refers to \textit{uniform transfers on the right} (UR). In this case, only the mean-preserving condition and the condition related to the right-hand side of the distribution (solidarity among the rich) need to be applied. By definition, an URL transfer is an UR transfer, but the converse is not true.
 
\begin{definition}[UR Transfer] \label{UR} Given two income distributions $\x, \y \in \ensn$ such that $\mu(\x)=\mu(\y)$, we say that $\x$ is obtained from $\y$ by means of a uniform-on-the-right progressive transfer, if there exist $\delta, \epsilon > 0$ and two individuals $1 \leq h<k \leq n$ such that $\x = \y + \xt$ and:
%
%
\begin{center}
\textnormal{
\small
\begin{tabular}{r*{11}{c}}
        \toprule
\textsc{Individual} =  & $1$  & \dots & $h-1$ & $h$       & $h+1$ & \dots & $k-1$ & $k$             & $k+1$            & \dots & $n$             \\
$\xt =$               & ($0$ & \dots & $0$    & $\delta$ & 0         & \dots & 0        & $-\epsilon$ & $-\epsilon$    & \dots & $-\epsilon$) \\
\bottomrule
\end{tabular}
}
\end{center}
\end{definition}

\vspace{0.2cm}
\noindent \textit{Uniform transfers on the left} (UL) are the symmetric counterpart of UR transfers. The mean-preserving condition is associated to the condition related to the left-hand side of the distribution (solidarity among the poor). Hence, an URL transfer is also an UL transfer (converse not true), but an UR transfer and an UL transfer are of different nature (neither is implied by the other).

\begin{definition}[UL Transfer] \label{UL} Given two income distributions $\x, \y \in \ensn$ such that $\mu(\x)=\mu(\y)$, we say that $\x$ is obtained from $\y$ by means of a uniform-on-the-left progressive transfer, if there exist $\delta, \epsilon > 0$ and two individuals $1 \leq h<k \leq n$ such that $\x = \y + \xt$ and:
%
%
\begin{center}
\textnormal{
\small
\begin{tabular}{r*{11}{c}}
        \toprule
\textsc{Individual} = & $1$         & \dots & $h-1$        & $h$       & $h+1$ & \dots & $k-1$ & $k$             & $k+1$ & \dots & $n$ \\
$\xt =$               & ($\delta$ & \dots & $\delta$    & $\delta$ & 0         & \dots & 0        & $-\epsilon$ & $0$     & \dots & $0$) \\
\bottomrule
\end{tabular}
}
\end{center}
\end{definition}

\vspace{0.2cm}
\cite{CM06} have established equivalence results comparable to Proposition~1, but by substituting the Pigou-Dalton transfers with the various alternatives presented here. First, they demonstrated that the utilitarian model is not flexible enough to distinguish between different principle of transfers. In each case, an equalising transfer (of whatever type) implies an increase in social welfare if and only if the utility function $u \in \ensU$ is concave. Hence, they obtain equivalence results between comparable statements (a), (b2) and (c), but not with (b1). For each equalising transfer (Definitions~\ref{URL}, \ref{UR} and~\ref{UL}), they identified the restriction to be placed on the weighting function $f \in \ensF$, and the corresponding implementation preorder (to be used instead of the Lorenz criterion).

The theory presented in this section can be used to replace the traditional theory based on the Pigou-Dalton principle of transfers. Nevertheless, the question of the relevance of the different definitions of  what we call an equalising transfer, is purely normative. The aim of this paper is to check whether these views on inequality are in line with people's preferences. 

\subsection{An overview of some existing experimental results}
\label{sec-expe-lit}

The Pigou-Dalton principle of transfers has already been explored in the experimental literature \cite[initiated by][]{AC92}. When the questions asked to the respondents are pairwise comparisons of income distributions (one before transfer, the other after, as in Figure~\ref{Trans-Illustr}), asking which is less unequal, the acceptance rates are generally little more than 50\%. At the opposite, the alternative principles of transfers, as presented in the previous section, have never been directly tested. We report here the results of several papers in the same series, in which the respondents were asked the same list of numerical questions (with some nuances between papers, mainly on the description of the context presented to the subject). Even if this was not the direct objective of these studies, some questions were compatible with the different principles. These papers are:  \cite{AC02}, \cite{GN03}, \cite{ACS04}, \cite{ACG09} and \cite{ACG12}.

The common questions asked to the respondents are presented in Table~\ref{expe-results-a1}. The initial distribution, denoted A, represents the income of 5 individuals, presented as identical in every respect other than their incomes. Incomes range from 2 units to 30. We also note that distribution B is always obtained from distribution A by a Pigou-Dalton transfer. 
\begin{table}[!htp] \caption{\label{expe-results-a1} \textit{The questionnaire}}
\vspace{0,2cm}
    \centering
    \scriptsize
    \begin{tabular}{*{11}{c}} 
    \toprule
    \multirow{2}{*}{\textsc{Questions}} & \multicolumn{10}{c}{Is the inequality higher in A or in B?} \\
    \cmidrule(lr){2-11}
     & \multicolumn{5}{c}{\textsc{Distribution A}} & \multicolumn{5}{c}{\textsc{Distribution B}} \\
    \cmidrule(lr){1-1} \cmidrule(lr){2-6} \cmidrule(lr){7-11}    
    \textbf{Q1}                                              & 2   & 5   & 9   & 20 & 30 & 2   & 6   & 8   & 20 & 30\\
    \textbf{Q2}                                              & 2   & 5   & 9   & 20 & 30 & 3   & 5   & 9   & 20 & 29\\
    \textbf{Q3}                                              & 2   & 5   & 9   & 20 & 30 & 2   & 6   & 9   & 20 & 29\\
    \hspace{0.15cm}\textbf{Q4}$^{\ast}$  & 2   & 5   & 9   & 20 & 30 & 2   & 10 & 9   & 15 & 30\\
    \textbf{Q5}                                              & 10 & 10 & 10 & 10 & 30 & 10 & 10 & 10 & 20 & 20\\
    \textbf{Q6}                                              & 2   & 5   & 9   & 20 & 30 & 2   & 6   & 9   & 19 & 30\\
    \bottomrule \\
    \multicolumn{11}{l}{$\ast$: Distribution B does not preserve the ranks of distribution A.}\\
    \end{tabular}
\end{table}
The questionnaire, which included other questions depending on the specific theme of each experiment, was distributed to a group of students at different universities. The particularities of each experiment are briefly presented in Table~\ref{expe-results}.

\begin{table}[!htp] \caption{\label{expe-results} \textit{Details on the experimental studies}}
\vspace{0,2cm}
    \centering
    \scriptsize
    \begin{tabular}{*{4}{l}} 
    \toprule
    \textsc{Papers} & \textsc{Year of the expe.} & \textsc{Subjects} (country \& number) & \textsc{Comments} \\ 
    \cmidrule(lr){1-1} \cmidrule(lr){2-2} \cmidrule(lr){3-3} \cmidrule(lr){4-4}
    A \& C (2002) & 1998/1999 & 587 students, 7 countries & Comparison inequality vs. risk perceptions \\
    G \& N (2003) & 1999/2000 & 159 students from Germany & Inequality vs. risk, plus `income gap' perceptions \\
    A, C \& S (2004) & 1994 & 510 students, 17 univ. in USA & Impact of socio-demo characteristics \\
    A, C \& G (2009) & from 2003 to 2006 & 653 students, Germany, Israel, UK & Position (involved or not) of the respondent \\
    A, C \& G (2012) & 2003 & 134 students, Germany, Israel, UK & 7 presentations for the same distributional problem \\
    \bottomrule 
    \end{tabular}
\end{table}

The results are reported in Table~\ref{expe-results-a2}. We also indicate the types of transfer with which each question is compatible. 
\begin{table}[!htp] \caption{\label{expe-results-a2} \textit{The results for all the experiments}}
\vspace{0,2cm}
    \centering
    \scriptsize
    \begin{tabular}{l *{7}{c}} 
    \toprule
    & \multicolumn{6}{c}{\textsc{Percentage of answer A}$^{\ast\ast}$} & \multirow{2}{*}{\textsc{Answer A to}}\\
    \cmidrule(lr){2-7}
    \textsc{Questions}$^{\ast}$            & \textbf{Q2} & \textbf{Q5} & \textbf{Q3} & \textbf{Q4} & \textbf{Q6} & \textbf{Q1} & \textsc{all questions} \\
    \textsc{Transfers}                           & URL            & UR          & UR           & PT              & PT               & PT              &                    \\
    \cmidrule(lr){1-1}  \cmidrule(lr){2-7} \cmidrule(lr){8-8}
    Amiel, and Cowell (2002)              & 74\%          & 72\%          & 61\%          & 60\%          & 48\%          & 40\%         & 17\%           \\
    Gaernter and Namazie (2003)       & 63\%          & 70\%          & 57\%          & 54\%          & 40\%          & 33\%         & 13\%           \\
    Amiel, Cowell and Slottje (2004)    & 54\%          & 54\%          & 47\%          & 45\%          & Not tested   & 34\%         & 10\%           \\
    Amiel, Cowell and Gaertner(2009)  & 80\%          & 77\%          & 71\%          & 61\%          & 58\%          & 58\%         & 26\%           \\
    Amiel, Cowell and Gaertner (2012) & 80\%          & 75\%          & 59\%          & 57\%          & 44\%          & 36\%         & 13\%           \\
    \bottomrule \\
    \multicolumn{8}{l}{\hspace{0.05cm} $\ast$: Questions are ordered according to decreasing acceptance rates in Amiel and Cowell (2002).}\\
    \multicolumn{8}{l}{$\ast\ast$: For all studies but A, C \& G (2009), we only report the results of the inequality questionnaire;}\\
    \multicolumn{8}{l}{\hspace{0.4cm} For A, C \& G (2009), we only report the results of the ``respondent as external observer'' scenario.}\\
    \end{tabular}
\end{table}
Only UL transfers are not represented. Then, we know that URL, UR and PT transfers are not independent. When PT is written in the table, it refers to a transfer that is not of a UR or URL type (hence, without any restriction on the right or on the left). Similarly, UR is not a URL type transfer. We can immediately see that the acceptance rates vary greatly between the questions, in an order that is compatible with the ethical requirements described above: On average, URL transfers are considered to be more equalising than UR transfers, and UR transfers more than (standard) PT transfers. We can conclude that the distinctions made between the different types of transfer in the previous section are echoed in the population of respondents (in this case, students). 

Strictly speaking, individuals with preferences compatible with the Pigou-Dalton principle of transfers should answer A to all the questions. We note that the percentage of subjects in this case is very low: it never exceeds a quarter of the population (last column in the table).  Even if uniform transfers were not formally defined and directly tested in these experiments, the intuition was already there. In \cite{AC02} it is written, pages 90 and 91: `\textit{Accordingly it may also be useful to consider a weaker version of same idea (\textnormal{namely, the principle of transfers}) that allows for the possibility that more complex criteria could be applied by individuals to inequality or risk comparison. An obvious example would be this criterion \dots If ceteris paribus a small amount of income is transferred from the person with the lowest income to the person with the highest income inequality must rise}'.

\section{Experimental design}
\label{sec-design}

\noindent \textsc{An experiment in three parts}. The experiment was divided into three parts. In the first part, the respondents were presented with a list of numerical questions, in which they had to compare a pair of distributions A and B. They were asked to indicate which distribution they thought was less unequal. A brief justification of the existence of the distributions was proposed in the instructions. The questions were presented one by one on the screen. In the second part, the acceptability of the uniform transfers was tested on the basis of text-based questions. The last part was a socio-demographic questionnaire. Detailed instructions can be found in Appendix~\ref{ap-instructions}. 

\noindent \textsc{Construction of distribution pairs}. The pairs of distributions under comparison were constructed as follows. First, we considered 5 initial distributions, denoted  $\y^1$,  $\y^2$,  $\y^3$,  $\y^4$ and  $\y^5$ in Table~\ref{Distrib}.
\begin{table}[!htp] \caption{\textit{Initial distributions}} \label{Distrib}
    \vspace{0.2cm}
    \centering
    \footnotesize
    \begin{tabular}{*{10}{c}}
        \toprule
         & \multicolumn{8}{c}{\textsc{Income scale}} \\
         \cmidrule(lr){2-10}
        \textsc{Distrib.}        & 2 & 4 & 6 & 8 & 10 & 12 & 14 & 16 & 18 \\
        \cmidrule(lr){1-1}                        \cmidrule(lr){2-10}
        $\y^1$                              & 2 & - & 6 & - & 10 & -    &14  & -    & 18  \\
        $\y^2$                              & 2 & 4 & -  & - & -    & -    & 14 & 16 & 18 \\
        $\y^3$                              & 2 & 4 & 6 & -  & -   & -   & -    & 16 & 18 \\
        $\y^4$                            & 2 & - & - & 8 & 10 & 12 & -    & -    & 18 \\
        $\y^5$                            & 2 & 4 & - & - & 10 & -    & -    & 16 & 18 \\
        \bottomrule
    \end{tabular}
\end{table}
Each distribution is an income list for 5 individuals, presented as perfectly identical apart from their income. The income scale was between 2 and 18 income units. These 5 distributions were designed to consider different distribution profiles, with a uniform distribution ($\y^1$) and different unimodal, skewed, and/or polarised distributions ($\y^2$, $\y^3$, $\y^4$ and $\y^5$). The objective was to find out whether the initial distributional structure could affect the acceptance of the various transfers.

We then considered all possible transfers of each type, limited to transfers between two individuals and of a single unit of income, as detailed in Table~\ref{Transfo}. 
\begin{table}[!htp] \caption{\textit{Equalising transfers}} \label{Transfo}
    \vspace{0.2cm}
    \centering
    \footnotesize
    \begin{tabular}{ll *{5}{c}}
        \toprule
        \multicolumn{2}{c}{\textsc{Transfers}} &  $e_1$ & $e_2$ & $e_3$ & $e_4$ & $e_5$ \\
        \cmidrule(lr){1-2}                    \cmidrule(lr){3-7}                    
        $\xt^1$ & URL                 			          & +1   &  0        & 0          & 0         & -1       \\
         \cmidrule(lr){1-2}                    \cmidrule(lr){3-7} 
        $\xt^2$ & UR                             		  & 0     &  +1      & 0          & 0         & -1       \\
        $\xt^3$ & UR                             		  & 0     &  0        &  +1       & 0         & -1       \\
        $\xt^4$ & UR                             		  & 0     &  0        & 0          &  +1      & -1       \\
        \cmidrule(lr){1-2}                    \cmidrule(lr){3-7} 
        $\xt^5$ & UL	                         & +1   &  0         & 0         & -1        & 0        \\
        $\xt^6$ & UL	                         & +1   &  0         & -1        &  0         & 0        \\
        $\xt^7$ & UL	                         & +1   &  -1        & 0         &  0         & 0        \\
        \cmidrule(lr){1-2}                    \cmidrule(lr){3-7} 
        $\xt^8$ & PT	                         & 0      & 0         & +1       & -1        & 0         \\
        $\xt^9$ & PT	                         & 0      & +1       & -1        & 0         & 0         \\
        $\xt^{10}$ & PT	                         & 0      & +1       & 0         & -1        & 0         \\
        \bottomrule
    \end{tabular}
\end{table}
\textit{URL transfers} indicates uniform-on-the-right-and-left transfers. \textit{UR transfers} indicate uniform-on-the-right transfers, which are not uniform-on-the-left. \textit{UL transfers} are defined symmetrically. \textit{PT transfers} are progressive transfers that are neither uniform on the right, nor on the left.\footnote{It is important to note that all the transfers considered here are, by definition, progressive transfers. In the same way, a URL transfer is also, by definition, a UR transfer or a UL transfer. In the analysis, we have organised the results in this way, in order to focus on the specific characteristics of each type of transfers.} We obtain one URL transfer and three transfers for all the other types (hence a total of 10 possible transfers).  Finally, we have considered all possible combinations of initial distributions and transfers, so that the final distribution can be written as $\x^i = \y^j + \xt^k$.

For each initial distribution, we have also added a \textit{test question}. In each case, the final distribution was the perfectly egalitarian distribution resulting from the initial distribution. For example, for $\y^1 = (2,6,10,14,18)$, the final distribution is $\x = (10,10,10,10,10)$. As control is non-existent in a web-experiment, we wanted to ensure that respondents did not answer completely randomly. In the case of the test questions, it seems obvious that the most egalitarian distribution is the one after transfer(s). We therefore used the results to these questions as a filter. The full list of distribution pairs is shown in Table~\ref{list_question-tab} in Appendix~\ref{liste_question}.

\noindent \textsc{Sequence of questions}. The initial distribution, before transfer, was always indicated as distribution A, and placed on the left of the screen. The idea was not to generate too much confusion among respondents, as numerical questions are already complex enough for some people.  In the same vein, the questions corresponding to the same initial distribution were presented in a single block. Including the 10 transfers and the test question, each block consisted of 11 questions. Within each block, the questions appeared on the screen (one per screen) in a random order. 

The block corresponding to the initial distribution $\y^1$ always appeared first. This distribution being uniform, the answers to the corresponding questions could not be `altered' by the degree of polarisation of this distribution. Also, as this block appeared first, the answers to the other blocks could not affect the answers to this one (controlling for a possible learning effect). Consequently, the acceptance rate of the transfers in this block can be considered as the best representation of respondents' preferences. 

The order of the other initial distributions was randomly selected. In order to limit the number of questions, we presented the respondent with only 4 of the 5 initial distributions: distributions $\y^1$, $\y^2$ and $\y^3$  and, by a 50/50 draw, either $\y^4$ or $\y^5$. In total, the respondent had to answer 44 questions. After each block of 11 questions, the respondent saw a screen, summarising their answers to that block. They could then modify their answers to the questions in that block.

\noindent \textsc{Text-based questions}. The preceding questions, which involve comparing lists of numbers, may prove complex for some participants. In order to corroborate or refute the results on these numerical questions, the 4 principles of transfers have also been presented to the participant in the form of text-based-questions, one per principle (see Appendix~\ref{ap-instructions}, Screen 4). The possible answers, for each question, were: Strongly disagree / Somewhat disagree / No opinion / Somewhat agree / Strongly agree. As with the numerical questions, to make sure that the participants had understood, we asked the following question at the end: `Did you find these questions clear?' The possible answers were: Not clear at all / Not clear / No opinion / Rather clear / Really clear. In analysing the results, we first look at the responses from the full sample, and then at those from participants who considered the questions to be (rather or really) clear.

\section{Sample of participants}
\label{sec-sample}

\noindent \textsc{Full sample of respondents}. The web-experiment was conducted in January 2021, with a sample of 1,028 respondents. The sample was representative of the French population (quota sampling method), taking into account the following selection criteria: gender, age (18 and over, including retired people), employment statuses and income. Sampling was carried out by a private company and the respondents were paid approximately \euro2 for their participation.\footnote{More details are available upon request. The experiment was made possible thanks to the financial support of the research project \textit{RediPref} (Contract ANR-15-CE26-0004).}
The fact of having a representative sample of a country's population is a specific feature of our paper that, to our knowledge, is not found in any other study on the same subject. 

The full sample of participants is described in Table~\ref{socio-demo-tab}, Appendix~\ref{ap-survey}. In addition to the criteria used to construct the sample, we also have socio-demographic information on the following variables: number of children, marital status, occupation category, level of education, whether or not they voted in the last presidential election, and their political opinions (right-left). Information we have but not included in the table are citizenship (97.67\% are French citizen) and department of residence (all French departments are represented, in comparable proportions). In order to check the representativeness of the sample, we have added a column indicating the percentages in each category, as computed by the official French statistical institute (INSEE). Note that all statistics are based on people aged 15 and over, so that some INSEE values have had to be recalculated (e.g. employment status, for which INSEE statistics are provided for individuals aged 15-64). Whatever the socio-demographic variable (including those not used for sampling), our full sample is very close to the national distribution. The main differences are as follows:
\begin{itemize}
\item[--] We have an under-representation of both the youngest and oldest age groups. The minimum age observed is 21, while the first age group defined by INSEE includes individuals aged 15-29. Additionally, recruiting older participants for studies of this kind is often challenging. In our sample, only one person fell into the oldest age group, being 94 years old at the time of the study.
\item[--] The difference in the number of children is due to the fact that INSEE only considers children under 25, whereas we consider all children. Thus, we have an under-representation for the category `no children'.
\item[--] The levels of education observed in the sample are quite different from national data. We have a very clear under-representation of the lowest level, namely ‘primary education’ (3\% in our sample, compared with 22\% nationally). At the opposite, the three highest levels are over-represented. There are at least two reasons for this. Firstly, level of education was not a sampling criterion. Secondly, such studies are conducted online, which implies that participants are familiar with digital tools. We can reasonably assume that they are less accessible to the less educated.
\item[--] We have a slight under-representation of the highest decile in gross monthly income. This is a frequent problem of access to high incomes through surveys. 
\end{itemize}
  
\noindent \textsc{Test-questions and restricted sample}. For both the numerical and text-based questions, we included test questions to check the respondents' understanding. For the 5 numerical questions (see Table~\ref{list_question-tab}), the more equal distribution was undoubtedly distribution B (10 income units for each). Any other answer was considered an error. On average, 61.75\% of the test questions were answered correctly (21.45\% for `Distribution A' and 16.80\% for `Neutrality'). This is a low percentage, which calls into question the respondents' overall understanding of our study. Our main concern here is that all respondents may have made at least one error. Fortunately, this is not the case.  Table~\ref{Tests} groups the subjects by the number of errors made. Remember that the subjects were confronted with only 4 of the 5 possible blocks of questions, which implies that the maximum number of errors is 4. We note that 43.58\% of them answered the 4 questions correctly. If we combine this result with the overall percentage of errors, this number is quite satisfactory.
\begin{table}[!htp] \caption{\textit{Number of errors on numerical test questions}} \label{Tests}
    \vspace{0.2cm}
    \centering
    \footnotesize
    \begin{tabular}{l *{5}{c}}
        \toprule
      						      & 4 \textsc{errors} & 3 \textsc{errors} & 2 \textsc{errors}  & 1 \textsc{errors} & 0 \textsc{error}  \\
        \cmidrule(lr){1-1}                 \cmidrule(lr){2-6}                                                   
        Number of subjects        &  205                     & 136                       & 106                       & 133                      & \textbf{448}                   \\
        Percentage of subjects & 19.94$\%$          & 13.23$\%$            & 10.31$\%$            & 12.94$\%$           & \textbf{43.58$\%$}          \\
        \bottomrule
    \end{tabular}\\
    \vspace{0.3cm}
\end{table}

In Table~\ref{socio-demo-tab}, where the socio-demographic characteristics of the sample are described, we have added a column entitled ‘restricted sample’, taking into account only respondents who made no errors in the test questions. Globally, the representativeness of the restricted sample, is not called into question: The percentage of respondents in each category rarely varies by more than 4\% compared to the full sample. The most notable change concerns the level of education: the restricted sample accentuates the under-representation of lower education levels, favoring higher levels (a result we could have anticipated).

In the analysis of the results, we first present those related to the full sample, followed by those concerning the restricted sample. The econometric analyses are focused on the restricted sample. This is an arbitrary choice. The aim here is not to exclude people on the basis of criteria such as mathematical reasoning ability. The problem is that the answers to the test questions are the only controls for this web-experiment. Unfortunately, among the respondents who made errors, we cannot distinguish those who had difficulty understanding the questions (whom we could have retained in the main sample) from those who answered randomly, or not seriously. To ensure that the responses given could be reasonably linked to the true preferences of the participants we chose to include, in the main analyses, only those who answered all the test questions correctly (448 persons). The only relative loss in terms of representativeness, compared to the overall French population, concerns the level of education. In our defense, we note that almost all comparable academic papers focus on samples of higher education students, hence the most educated individuals.

In Table~\ref{Tests-text}, we present the responses to the question "Did you find these questions clear?" asked at the end of the 4 text-based questions (Appendix~\ref{ap-instructions}, Screen 4).
\begin{table}[!htp] \caption{\textit{Number of errors on text-based test questions}} \label{Tests-text}
    \vspace{0.2cm}
    \centering
    \footnotesize
    \begin{tabular}{l *{5}{c}}
        \toprule
      					   & \textsc{Not clear} & \textsc{Not}    & \textsc{No}        & \textsc{Rather} & \textsc{Really}  \\
	      			           & \textsc{at all}         & \textsc{clear} & \textsc{opinion} & \textsc{clear}    & \textsc{clear}  \\
        \cmidrule(lr){1-1}                 \cmidrule(lr){2-6}                                                   
        Number of subjects        &  109                     & 289                       & 179                       & \textbf{377}                      & \textbf{74}                   \\
        Percentage of subjects & 10.60$\%$          & 28.11$\%$            & 17.41$\%$            & \textbf{36.67$\%$}           & \textbf{7.20$\%$}          \\
        \bottomrule
    \end{tabular}\\
    \vspace{0.3cm}
\end{table}
We note that 43.87\% of respondents found the questions clear (rather clear + really clear). This percentage is comparable to that of respondents who did not make any errors on the numerical test questions. However, as shown in the Figure~\ref{Plot-errors}, Appendix~\ref{ap-survey}, the correlation between the number of errors on the numerical test questions, and the level of understanding of the text-based questions, is almost null.

\section{Acceptance rates of the different principle of transfers}
\label{sec-acceptance}

\subsection{Numerical questions}
\label{subsec-numerical}

\noindent \textsc{Full sample}. First, we present in Table~\ref{Resultats-globaux} the results for the full sample (1,028 people). We distinguish between four types of transfer: URL transfers, UR transfers, UL transfers and the standard PT transfers. We recall that, in all the tables presented below,  a `PT transfer' indicates a progressive transfer that is neither uniform on the right nor on the left. An `UR transfer' is a uniform-on-the-right transfer, which is not uniform-on-the-left. Finally, an `UL transfer' indicates a uniform-on-the-left transfer, which is not uniform-on-the-right. 
\begin{table}[!htp] \caption{\textit{Acceptation rates for the numerical questions (all the subjects)}} \label{Resultats-globaux}
    \vspace{0.2cm}
    \centering
    \footnotesize
    \begin{tabular}{l *{3}{c}}
        \toprule
        \textsc{Transfers}  & \textsc{Accepted} & \textsc{Rejected} & \textsc{Neutrality}  \\
        \cmidrule(lr){1-1}                 \cmidrule(lr){2-4}                                                   
        URL      & 45.06$\%$           & 16.76$\%$           & 38.18$\%$                \\
        UL 	     & 41.67$\%$           & 19.33$\%$           & 39.00$\%$              \\
        UR        & 38.36$\%$           & 19.64$\%$           & 42.00$\%$                \\
        PT         & 31.75$\%$           & 22.05$\%$           & 46.20$\%$                 \\
        \cmidrule(lr){1-1}                 \cmidrule(lr){2-4}                                                  
        All transfers  & \textbf{38.04$\%$}           & 19.98$\%$                & 41.98$\%$              \\
        \bottomrule
    \end{tabular}\\
    \vspace{0.3cm}
\end{table}
We make a distinction between strict acceptance, neutrality and rejection. Strictly speaking, as transfers are generally defined in a weak sense, neutrality is consistent with the underlying principle of transfers. To avoid any ambiguity, we have nevertheless chosen to separate acceptance and neutrality. All transfers (uniform or not) being progressive transfers, the rate of acceptance of the principle of transfers is 38.04\%. This is relatively low compared to comparable studies, particularly those summarised in Table~\ref{expe-results-a2}. If we focus on non-uniform progressive transfers (PT transfers in the table), the acceptance rate is significantly lower (31.75\%). This indicates that transfers involving neither the poorest nor the richest are perceived as the most ambiguous in terms of reducing inequality. Such a low rate can be partly explained by the results obtained for the test questions recalling that, on average, respondents answered 61.75\% of these questions correctly.

\noindent \textsc{No errors in the test questions}. The results for the restricted sample of 448 people are presented in Table~\ref{Resultats-globaux-sans-erreur}. We note that the overall rate of acceptance of the principle of transfers has increased considerably (51.88\%), which makes it more comparable to that of the studies in Table~\ref{expe-results-a2}.  Despite this, acceptance of non-uniform progressive transfers (PT) remains low, at less than 40\%. 
\begin{table}[!htp] \caption{\textit{Acceptation rates for the numerical questions (no errors in the test questions)}} \label{Resultats-globaux-sans-erreur}
    \vspace{0.2cm}
    \centering
    \footnotesize
    \begin{tabular}{l *{3}{c}}
        \toprule
        \textsc{Transfers}   & \textsc{Accepted} & \textsc{Rejected} & \textsc{Neutrality}  \\
        \cmidrule(lr){1-1}                 \cmidrule(lr){2-4}                                                                                     
        URL      & 66.80$\%$           & 8.43$\%$              & 24.78$\%$                \\
        UL 	     & 59.30$\%$           & 12.02$\%$           & 28.68$\%$              \\
        UR        & 51.97$\%$           & 14.08$\%$            & 33.95$\%$                \\
        PT         & 39.38$\%$           & 18.47$\%$           & 42.15$\%$                 \\
        \cmidrule(lr){1-1}                 \cmidrule(lr){2-4}                                                  
        All transfers  & \textbf{51.88$\%$}           & 14.21$\%$          & 33.91$\%$              \\
        \bottomrule
    \end{tabular}\\
    \vspace{0.3cm}
\end{table}
Overall, acceptance rates differ between the types of transfer. First, uniform transfers are much widely accepted than non-uniform transfers. By far, the most widely accepted transfers are those that combine uniformity on the right and left (URL). While this result seems fairly intuitive, the comparison of UR and UL transfers is not. We find that respondents perceive uniform transfers to the left as having a greater capacity to reduce inequalities. This is an important result of our analysis: when a reduction in inequality is suitable, it seems preferable to give priority to reducing poverty, rather than reducing the gap with the rich. All the differences between the acceptance rates are significant ($\chi^{2}$ statistics), as shown in Table~\ref{Equality-global}.

\begin{table}[!htp]\caption{\textit{Equality tests of the acceptance rates on the numerical questions}} \label{Equality-global}
    \vspace{0.2cm}
    \centering
    \footnotesize
\begin{tabular}{lccc}
\toprule
$\chi^{2}$ \textsc{Statistics} & \textsc{DL} & \textsc{Value} & \textsc{Prob.} \\
	\cmidrule(lr){1-1} \cmidrule(lr){2-2} \cmidrule(lr){3-4}
Global& 3 & 614.86 & < 0.0001 \\
URL versus UR& 1 & 119.70 & < 0.0001 \\
URL versus UL & 1 & 31.80 & < 0.0001 \\
URL versus P& 1 & 406.45 & < 0.0001 \\
UR versus UL& 1 & 58.49 & < 0.0001\\
UR versus P& 1 & 171.79 & < 0.0001 \\
UL versus P& 1 & 426.80 & < 0.0001 \\
\bottomrule
\end{tabular}\\
    \vspace{0.2cm}
    \textit{Notes}. Null hypothesis $\rightarrow$ equality of the acceptance rates.
\end{table}

\noindent \textsc{Impact of initial distributions}. In Table~\ref{Resultats-distributions}, we distinguish the results for the five initial distributions  (before transfers). The aim was to see whether the acceptability of the various transfers could be affected by the structure, more or less polarised, of the initial distribution. 
\begin{table}[!htp] \caption{\textit{Acceptation rates by initial distribution}} \label{Resultats-distributions}
    \vspace{0.2cm}
    \centering
    \footnotesize
    \begin{tabular}{l *{5}{c}}
        \toprule
        \textsc{Transfers}   & $\y^1$  & $\y^2$ & $\y^3$ & $\y^4$ & $\y^5$   \\
        \cmidrule(lr){1-1}                 \cmidrule(lr){2-6}                                                                                     
        URL        & 66.96$\%$   & 66.74$\%$   & 66.29$\%$    & 67.42$\%$   & 66.96$\%$ \\
        UL 	      & 58.04$\%$   & 60.71$\%$   & 57.89$\%$    & 60.48$\%$   & 60.65$\%$ \\
        UR         & 47.84$\%$   & 52.68$\%$   & 54.84$\%$    & 51.58$\%$   & 53.45$\%$ \\
        PT          & 31.62$\%$   & 42.71$\%$   & 41.29$\%$    & 42.08$\%$   & 41.70$\%$ \\
        \cmidrule(lr){1-1}                 \cmidrule(lr){2-6}                                                  
        All transfers  & 47.95$\%$   & 53.50$\%$   & 52.83$\%$    & 52.99$\%$   & 53.44$\%$ \\
        \bottomrule
    \end{tabular}\\
    \vspace{0.3cm}
\end{table}
No clear trend emerges from this table. The results are fairly similar to those in Table~\ref{Resultats-globaux-sans-erreur}. We simply note that the lowest acceptance rates concern the uniform initial distribution. The fact that the initial distribution has no impact on acceptability is consistent with the theory, which always defines a transfer that reduces inequality, independently of the initial distribution.

\noindent \textsc{Gender}. The influence of socio-economic variables on the results is presented in the various tables in Appendix~\ref{other-tables}. The results by gender are shown in Table~\ref{Resulats-genre}. The first observation is that the ranking of acceptance rates URL > UL > UR > PT is found for both men and women (the differences are significant, see Table~\ref{Equality-gender}). Another interesting result is that, whatever the type of transfer, apart from non-uniform transfers (PT), the acceptance rate is always significantly higher for men than for women (see Table~\ref{Equality-gender2}). 

\noindent \textsc{level of education}. In Tables~\ref{Resulats-ecole} to~\ref{Equality-schooling2}, we distinguish results by level of education. A first signal is the low acceptance rate for people with a `before high school' level of education, for all transfers. One possible explanation is that these people may have difficulty understanding the questions, as they may not be at ease with the mathematical formalism. Such an interpretation seems plausible, especially if we focus on URL transfers. Indeed, it is difficult to consider that such transfers (involving only the richest and the poorest in this study), are not perceived as reducing inequality. We note here that the acceptance rate increases significantly with the degree (ranging from 45.83\% to 71.18\%). For UR and URL transfers, we also note an increase in acceptance with the degree, except between the last two levels (`short tertiary education' and `university degree'). On the other hand, for PT transfers, although the differences are significant (Table~\ref{Equality-schooling2}), the differences are smaller. This confirms the fact that non-uniform transfers are perceived as having an ambiguous effect on overall inequality. Finally, in Table~\ref{Equality-schooling}, we note that the ranking URL > UL > UR > PT is significant whatever the level of education, except for `before high school'. 

\noindent \textsc{Political opinions}. Political opinions also have a significant influence on results. Whatever the type of transfer, the acceptance rates are significantly different (see Table~\ref{Equality-politique2}). If we look in detail, the ranking URL > \dots > PT is again found, for all political opinions (sometimes non-significant differences, see Table~\ref{Equality-politique}). A first strong result is a decrease in the acceptance of URL transfers, as we move from `far left' opinions to `far right' opinions (Table~\ref{Resulats-politique}). Another interesting result is the comparison of UR and UL transfers. For UR transfers, which include a form of solidarity among the rich (who are donors) but which do not involve solidarity among the poor (who are receivers), acceptance is higher among `far right' people than among `far left' people (54.58\% vs. 47.78\%). On the other hand, for UL transfers (which imply solidarity in the other direction), the result is reversed (65.00\% for `far left' vs. 55.83\% for `far right'). 

\noindent \textsc{Professional status}. The latest analyses, in Tables~\ref{Resulats-profession} to~\ref{Equality-profession2}, focus on professional status. Here, the results are much more ambiguous. For instance, for many statuses, the difference between URL and UL, or between UR and UL, is not significant. The only really striking result is the greater acceptance of all transfers, except PT, by 'part-timers' compared with 'full-timers'. 

\subsection{Text-based questions}
\label{subsec-text}

The text-based questions are presented in Appendix~\ref{ap-instructions}, Screen 4. The results for the full sample of participants are centralised in Table~\ref{Resultats-globaux-text}.
\begin{table}[!htp] \caption{\textit{Acceptation rates for the text-based questions (all the subjects)}} \label{Resultats-globaux-text}
    \vspace{0.2cm}
    \centering
    \footnotesize
    \begin{tabular}{l *{3}{c}}
        \toprule
        \textsc{Transfers}  & \textsc{Accepted} & \textsc{Rejected} & \textsc{Neutrality}  \\
        \cmidrule(lr){1-1}                 \cmidrule(lr){2-4}                                                   
        URL      & 50.78$\%$           & 20.53$\%$           & 28.70$\%$                \\
        UL 	     & 47.76$\%$           & 29.28$\%$           & 22.96$\%$              \\
        UR        & 49.12$\%$           & 26.17$\%$           & 24.71$\%$                \\
        PT         & 46.30$\%$           & 35.12$\%$           & 18.58$\%$                 \\
        \cmidrule(lr){1-1}                 \cmidrule(lr){2-4}                                                  
        All transfers  & \textbf{48.49$\%$}           & 27.77$\%$                & 23.74$\%$              \\
        \bottomrule
    \end{tabular}\\
    \vspace{0.3cm}
\end{table}
As with the numerical questions, a majority of participants rejected the principle of transfers: this is validated in only 48.49\% of the responses. The hierarchy of acceptance rates between PT, UR, UL and URL observed in the numerical questions is confirmed, with a slight reverse for UR and UL (although the difference is very small). Generally speaking, the difference between the acceptance rates is much smaller (only a difference of 4\% between PT and URL). The acceptance rates for the sample restricted to participants who found the questions clear, is presented in Table~\ref{Resultats-globaux-sans-erreur-text}. 
\begin{table}[!htp] \caption{\textit{Acceptation rates for the text-based questions (questions perceived as clear)}} \label{Resultats-globaux-sans-erreur-text}
    \vspace{0.2cm}
    \centering
    \footnotesize
    \begin{tabular}{l *{3}{c}}
        \toprule
        \textsc{Transfers}   & \textsc{Accepted} & \textsc{Rejected} & \textsc{Neutrality}  \\
        \cmidrule(lr){1-1}                 \cmidrule(lr){2-4}                                                                                     
        URL      & 68.96$\%$           & 19.07$\%$              & 11.97$\%$                \\
        UL 	     & 62.08$\%$           & 28.16$\%$           & 9.76$\%$              \\
        UR        & 64.30$\%$           & 23.50$\%$            & 12.20$\%$                \\
        PT         & 57.21$\%$           & 34.81$\%$           & 7.98$\%$                 \\
        \cmidrule(lr){1-1}                 \cmidrule(lr){2-4}                                                  
        All transfers  & \textbf{63.14$\%$}          & 26.38$\%$          & 10.48$\%$              \\
        \bottomrule
    \end{tabular}\\
    \vspace{0.3cm}
\end{table}
It confirms the hierarchy between the different principles, with always a very small gap between the different rates. However, we note that the acceptance rates increase sharply compared to the full sample. In Table~\ref{Equality-global-text}, we test the significance of the differences between the rates. As these rates are very close, many of the differences are not significant. This is particularly the case for the UR versus UL comparison. The dominance of UL over UR, as observed in the numerical questions, is therefore not called into question here. 

\begin{table}[!htp]\caption{\textit{Equality tests of the acceptance rates on the text-based questions}} \label{Equality-global-text}
    \vspace{0.2cm}
    \centering
    \footnotesize
\begin{tabular}{lccc}
\toprule
$\chi^{2}$ \textsc{Statistics} & \textsc{DL} & \textsc{Value} & \textsc{Prob.} \\
	\cmidrule(lr){1-1} \cmidrule(lr){2-2} \cmidrule(lr){3-4}
Global& 3 & 13.86 & 0.003 \\
URL versus UR& 1 & 2.20 & \textbf{0.138} \\
URL versus UL& 1 & 4.72 & 0.030 \\
URL versus PT& 1 & 13.37 & < 0.0001 \\
UR versus UL& 1 & 0.48 & \textbf{0.490}\\
UR versus PT& 1 & 4.76 & 0.029 \\
UL versus PT& 1 & 2.23 & \textbf{0.135} \\
\bottomrule
\end{tabular}\\
    \vspace{0.2cm}
    \textit{Notes}. Null hypothesis $\rightarrow$ equality of the acceptance rates.
\end{table}

\section{Econometric estimations of the social welfare functions}
\label{sec-sw}

\subsection{Econometric strategy}
\label{subsec-ecok}

The theory of inequality measurement assumes that individual preferences can be represented by a social welfare function. The two models considered in this paper are the utilitarian approach $W_u$ and the extended Gini approach $W_f$ (see Section~\ref{sec-framework}). To the best of our knowledge, attempts to estimate the utility model have been proposed by \cite{ACH99} and \cite{CDJS05}, but no paper has studied the extended Gini approach. The paper closest to ours, in terms of methodology, is \cite{HO94}'s, but applied to individual decisions under risk. 

In each of the 40 numerical questions (test questions are excluded), the respondent has to compare two distributions $\x$ and $\y$. By indicating which distribution is considered as more equal, the respondent provides an indication on $\Delta_W(\x,\y; \alpha)$, defined as follows:
\begin{equation} 
\Delta_W(\x,\y; \alpha) = \alpha \left[ W(\x) - W(\y) \right]\,,
\end{equation}
with $\alpha > 0$ a free parameter. If distribution $\x$ (resp. $\y$) is strictly preferred, then $\Delta_W(\x,\y; \alpha) > 0$ (resp. $<0$). If the level of inequality is considered to be the same in both distributions, then $\Delta_W(\x,\y; \alpha) = 0$. Although preferences are assumed to be deterministic (and representable by a social welfare function), some errors are possible when the respondent answers the questions. To this end we add a white noise, normally distributed. We obtain a stochastic specification for the estimation model:
\begin{equation} \nonumber
\Delta_W^{\star}(\x,\y; \alpha) \defn \Delta_W(\x,\y; \alpha) + \varepsilon\,, \qquad
\textnormal{where}\ \varepsilon \sim N(0;1)\,.
\end{equation}

\noindent Whereas $\Delta_W(\x,\y; \alpha)$ is positive if the distribution $\x$ is \textit{preferred} by the respondent, we only observe $\Delta_W^{\star}(\x,\y; \alpha)$, which is positive if $\x$ is \textit{chosen}. Then, given that only the ordinal information of $\Delta$ is meaningful here (the intensity of the difference in social welfare cannot be interpreted) we replace, for econometric estimation purposes, $\Delta$ by a discrete variable $\gamma$, such as:
\begin{displaymath}
\left\{ \begin{array}{ll}
\gamma=0 & \textrm{if}\quad \Delta_W^{\star} < \tau_1\,,\\
\gamma=1 & \textrm{if}\quad \tau_1 \leq \Delta_W^{\star} \leq \tau_2\,,\\
\gamma=2 & \textrm{if}\quad \Delta_W^{\star} > \tau_2\,.
\end{array} \right.
\end{displaymath}
The threshold parameters $\tau_1$ and $\tau_2$ have to be estimated, with $\tau_1 \leq 0$ and $\tau_2 \geq 0$. We obtain an \textit{ordered probit model}, estimated by applying \textit{maximum log-likelihood methods}.\footnote{We use the \textit{maxLik function} in the library of the same name, available in the freeware R.} 

As different optimisation algorithms can lead to different results, we propose to apply two approaches. First, we apply a quasi-Newton method, called \textit{Broyden-Fletcher-Goldfarb-Shanno} (BFGS). The algorithm iteratively updates parameter estimates by considering both the gradient of the objective function (in our case, the opposite of the log-likelihood function, which is minimised) and an approximation of the inverse Hessian matrix. Then, we apply the \textit{Simulated Annealing} (SANN) algorithm. In each iteration, the algorithm considers moves to both better and occasionally worse solutions, allowing it to escape local optima and explore a broader range of potential solutions. The likelihood of accepting a worse solution decreases over time, which gradually refines the search around the most promising parameter values.

\subsection{Parametric estimations}
\label{subsec-param_estim}

The \textit{Atkinson-Kolm-Sen class of inequality indices}  \cite[][]{K69, At70, Se73} is derived from the utilitarian social welfare function $W_u$, with the following utility function:
\begin{equation}\label{u_H}
    u_{\epsilon}(x_i) = \left\{
        \begin{array}{lll}
            \frac{1}{\epsilon}\,x_i^{\epsilon}\,, & \textrm{if}\quad \rho \neq 0\,, \\
            \ln x_i \,,                    & \textrm{if}\quad \rho = 0\,,
        \end{array} \right.
\end{equation}
where $\epsilon \leq 1$ is the inequality aversion parameter: the lower it is, the greater the aversion ($\epsilon=1$ indicates neutrality to inequality). This utility function is by definition concave, hence consistent with the Pigou-Dalton principle of transfers. The estimation is based on:
\begin{equation} 
\Delta_{{\epsilon}} (\x,\y; \alpha) = \frac{\alpha}{5} \sum_{i=1}^5 \left[ u_\epsilon(x_i) - u_\epsilon(y_i) \right]\,,
\end{equation}
where $\alpha$ and $\epsilon$ are the parameters to be estimated. Whereas $\alpha$ is constrained to be positive, no restriction is placed on $\epsilon$, so that $u_\epsilon$ can be concave, linear or convex. 

Alternatively, under the extended Gini approach, the \textit{Donaldson-Weymark class of inequality indices} \cite[][]{DW80} is derived from the social welfare function $W_f$, with the following weighting function: 
\begin{equation}\label{f_H}
    f_{\eta}(t) = t^\eta\,,\quad \eta \geq 1\,.
\end{equation}
The higher $\eta$ is, the greater the inequality aversion. This function, convex, is consistent with the principle of transfers. The Gini index is obtained with $\eta=2$. We first note that $W_f(\x) = \sum_{i=1}^{n} f (\frac{n-i+1}{n}) (x_{i} - x_{i-1})$. By letting  $d_i \defn (x_i - x_{i-1}) - (y_i - y_{i-1})$ and $x_0=y_0=0$, and observing that in our experiment $\mu(\x) = \mu(\y)$ in each question, the estimation is based on:
\begin{eqnarray}  
\Delta_{{\eta}} (\x,\y; \alpha) &=& \Delta_{{\eta}} (\x,\y; \alpha) - \alpha \left[ \mu(\x) - \mu(\y) \right]\,,  \\
 &=& \alpha \sum_{i=1}^{5} f_\eta \bigg(\frac{5-i+1}{5} \bigg) d_i - \alpha \sum_{i=1}^{5} \bigg(\frac{5-i+1}{5} \bigg) d_i\,,  \\
 &=& \alpha \sum_{i=2}^{5} \left[ f_\eta \bigg(\frac{5-i+1}{5} \bigg) - \bigg(\frac{5-i+1}{5} \bigg) \right] d_i\,. \label{param-f-full}
\end{eqnarray}
Again, $\alpha > 0$ a parameter to be estimated. We impose $\eta$ to be positive (to have $f \in \ensF$), and $f_\eta$ can be concave, linear or convex. 

The estimation results for the utility function $u_\epsilon$ and the weighting function $f_\eta$ are presented in Table~\ref{Resultats-shape}, for all the participants in the restricted sample.  
\begin{table}[!htp] \caption{\textit{Shapes of the utility function $u_\epsilon$ and the the weighting function $f_\eta$}} \label{Resultats-shape}
    \vspace{0.2cm}
    \centering
    \footnotesize
    \begin{tabular}{l *{6}{c}}
       \toprule
        & \multicolumn{3}{c}{\textsc{SANN Algorithm}} & \multicolumn{3}{c}{\textsc{BFGS Algorithm}} \\
       \cmidrule(lr){2-4} \cmidrule(lr){5-7}
       \textsc{Model}& Concave & Linear & Convex & Concave & Linear & Convex  \\
       \cmidrule(lr){1-1} \cmidrule(lr){2-4} \cmidrule(lr){5-7}                                                  
        Utilitarianism ($\epsilon$)  & \textbf{87.72}\% & -- & 12.28\% & \textbf{98.44}\% & -- & 1.56\% \vspace{0.2cm}\\ 
        Extended Gini ($\eta$) & 18.75\% & -- & \textbf{81.25}\% & 2.90\% & -- & \textbf{97.10}\% \\
        \bottomrule
    \end{tabular}\\
    \vspace{0.3cm}
\end{table}
As expected, whatever the optimisation method (SANN or BFGS), the preferences of a large majority of respondents are represented, respectively, by a concave utility function and a convex weighting function. This observation is all the more true for BFGS, where the rate is almost 100\% in both cases. We precise that convergence rate for the function $u_\epsilon$ is 100\% for SANN, and 77.90\% for BFGS. For the function $f_\eta$, it is 100\% for SANN, and 84.38\% for BFGS. The first method therefore seems more appropriate for estimating our models. 

An overview of the parameters for all participants is presented in Table~\ref{Resultats-Utility}. The mean and standard-deviation of the parameter $\epsilon$ are shown first, and then the median, the $1^{st}$ and $3^{rd}$ quartiles. 
\begin{table}[!htp] \caption{\textit{Estimation of $\epsilon$ for the utility function $u_\epsilon$}} \label{Resultats-Utility}
    \vspace{0.2cm}
    \centering
    \footnotesize
    \begin{tabular}{l *{2}{c}}
        \toprule
         & SANN \textsc{Algorithm} & BFGS \textsc{Algorithm} \\
       \cmidrule(lr){2-2}  \cmidrule(lr){3-3}    
        \textsc{Statistics}   & Parameter $\epsilon$  & Parameter $\epsilon$ \\
       \cmidrule(lr){1-1} \cmidrule(lr){2-2}  \cmidrule(lr){3-3}                                                                                         
        Mean                          & 0.67 & 0.43 \\
        Standard-Deviation    & 0.42 & 0.51 \\
        \cmidrule(lr){1-1} \cmidrule(lr){2-2}  \cmidrule(lr){3-3} 
        \textbf{Median}          & \textbf{0.69} & \textbf{0.26}  \\
        1st Quartile                & 0.34 & 0.07 \\
        3rd Quartile                & 0.84 & 0.76 \\      
        \bottomrule 
   \end{tabular}\\
    \vspace{0.3cm}
\end{table}
Due to the high variability of the parameter between individuals, the median is here a better indicator of the centre of distribution. The median parameter estimated with the SANN algorithm, equal to 0.69, is close to that of \cite{ACH99}, where a value of 0.75 is found. The median value estimated by BFGS is lower, indicating a stronger aversion to inequality. 
By taking as lower and upper bounds the union of, respectively, the $1^{st}$ and $3^{rd}$ quartiles for the two estimation methods, we find a range $\epsilon \in [0.07, 0.84]$, which is not so large. A Kernel density estimate of $\epsilon$ is provided in Figure~\ref{Kernel-U}, Appendix~\ref{app-kernel}. Parameter $\epsilon$ is clearly not normally distributed, each method finding a bimodal distribution, but with different shapes. 

\begin{figure}[!htp] \caption{\label{Param-estim} \textit{Median value of $\eta$ for weighting function $f_\eta$}} 
\vspace{0.2cm}
\small
\begin{tabular}{cc}
\includegraphics[width=8.5cm]{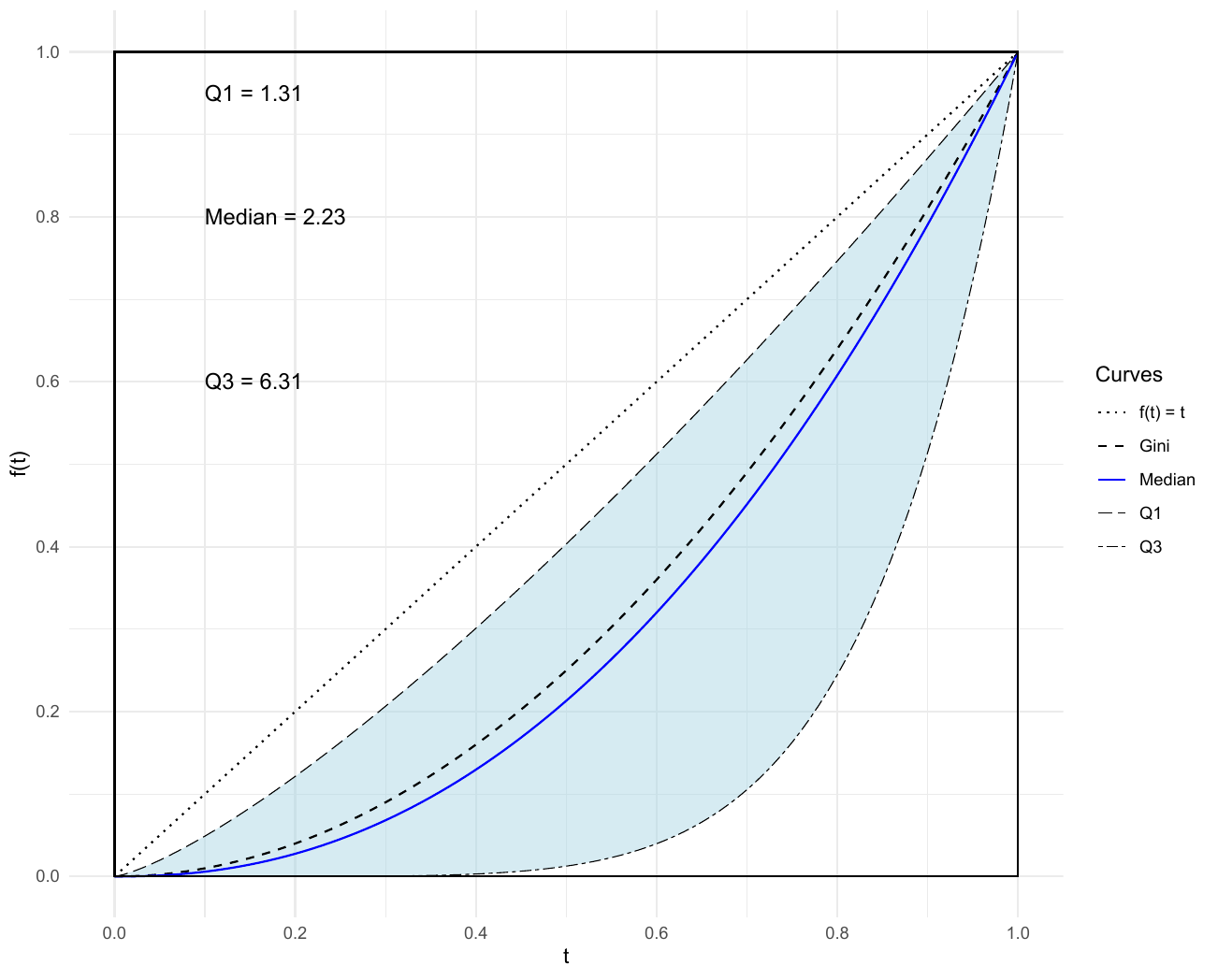} & %
\includegraphics[width=8.5cm]{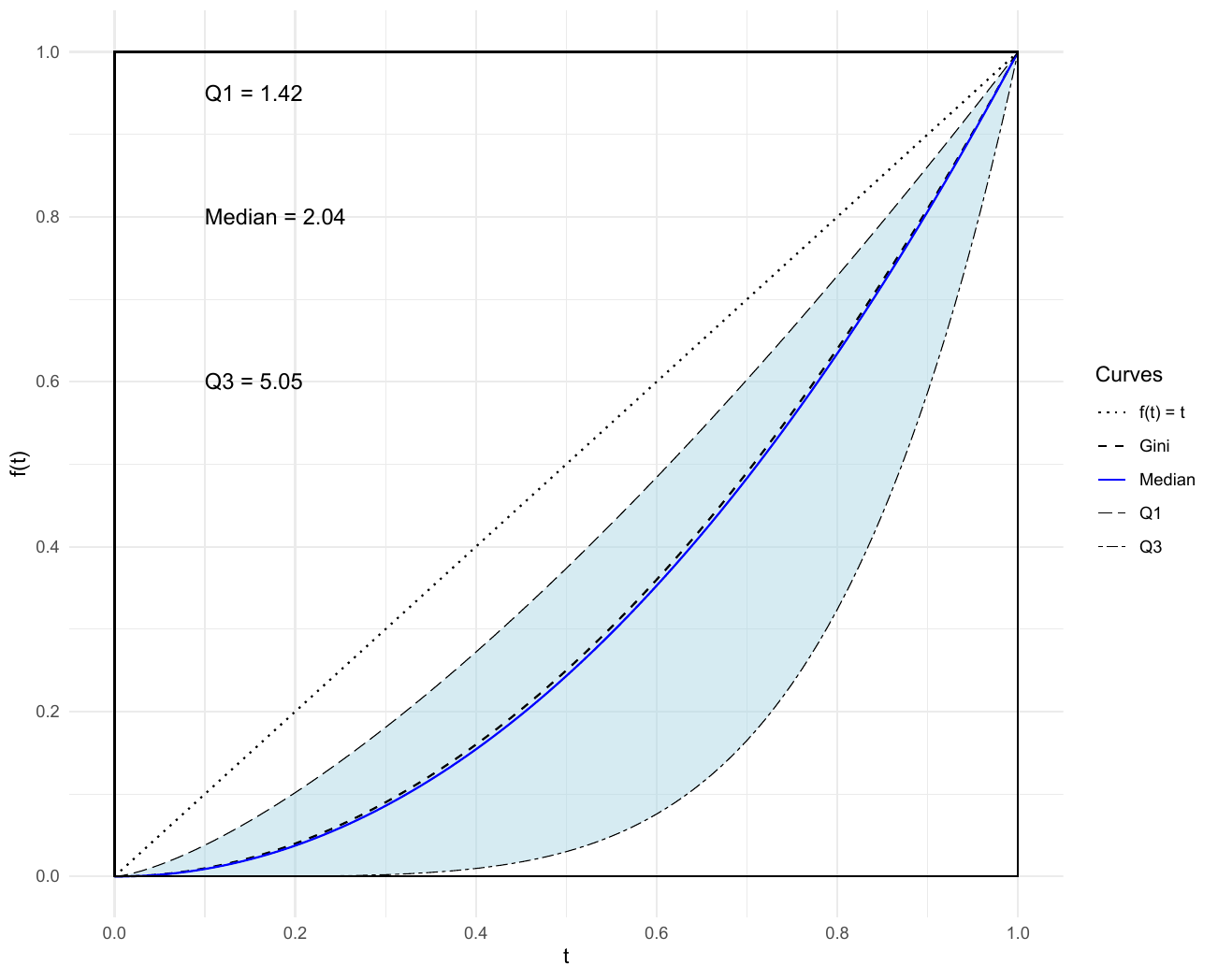} \\ 
\textit{SANN Algorithm} & \textit{BFGS Algorithm} 
\end{tabular}
\end{figure}

The same estimations are provided in Table~\ref{Resultats-extended-Gini} for the weighting function and the parameter~$\eta$. We also provide the difference between the estimated median value of $\eta$ and the value characterising the Gini index, $\eta=2$. 
\begin{table}[!htp] \caption{\textit{Estimation of $\eta$ for the weighting function $f_\eta$}} \label{Resultats-extended-Gini}
    \vspace{0.2cm}
    \centering
    \footnotesize
    \begin{tabular}{l *{2}{c}}
        \toprule
         & SANN \textsc{Algorithm} & BFGS \textsc{Algorithm} \\
       \cmidrule(lr){2-2}  \cmidrule(lr){3-3}    
        \textsc{Statistics}   & Parameter $\eta$  & Parameter $\eta$ \\
       \cmidrule(lr){1-1} \cmidrule(lr){2-2}  \cmidrule(lr){3-3}                                                                                         
        Mean                          &  4.90 & 7.38 \\
        Standard-Deviation    &  5.89 & 12.97 \\
        \cmidrule(lr){1-1} \cmidrule(lr){2-2}  \cmidrule(lr){3-3} 
        \textbf{Median}          & \textbf{2.23} & \textbf{2.04}  \\
        1st Quartile                & 1.31 & 1.42 \\
        3rd Quartile                & 6.31 & 5.05 \\      
        \cmidrule(lr){1-1} \cmidrule(lr){2-2}  \cmidrule(lr){3-3}  
        \textbf{Gini - Median} & \textbf{-0.23} & \textbf{-0.04} \\
        \bottomrule
    \end{tabular}\\
    \vspace{0.3cm}
\end{table}
We first note that the two algorithms converge towards a value for the median that is roughly equal. What's more, this value is almost identical to that for the Gini. This is a strong result which has never been observed, to our knowledge, in the literature. To confirm the closeness of our estimates, we have plotted the median value of $\eta$, as well as the $1^{st}$ and $3^{rd}$ quartiles (lower and upper bounds), against Gini in Figure~\ref{Param-estim}. 
The social welfare function underlying the Gini index is therefore a perfect proxy of the preferences of the median individual. A Kernel density estimate of $\eta$ is also provided in Figure~\ref{Kernel-F}, Appendix~\ref{app-kernel}. The different with the estimated density of $\epsilon$ is important. First, the distributions obtained with the two algorithms are similar, and unimodal. Both show a thick distribution tail on the right, suggesting the presence of individuals with a strong aversion to inequality. 

We then compare the parameter estimates of the two models, utilitarianism and Extended Gini. We note that the structure of these two models is theoretically similar, in the sense that there is only one main parameter to be estimated, that of a power function. The difference between the two approaches is that utilitarianism weights income levels, whereas Extended Gini weights rank in the distribution. We compare the models using the \textit{Akaike Information Criterion} (AIC). The results are presented in Table~\ref{Resultats-AIC}.
\begin{table}[!htp] \caption{\textit{Which parametric model best fits the data according to AIC?}} \label{Resultats-AIC}
    \vspace{0.2cm}
    \centering
    \footnotesize
    \begin{tabular}{l *{2}{c}}
        \toprule
      	\textsc{Model} & \textsc{SANN Algorithm} & \textsc{BFGS Algorithm}  \\
       \cmidrule(lr){1-1} \cmidrule(lr){2-3}                                                   
        Utilitarianism ($\epsilon$)   & 19.64\% & 17.63\% \\
        Extended Gini ($\eta$)  & \textbf{71.43}\% & \textbf{68.08}\% \\
        \bottomrule
    \end{tabular}\\
    \vspace{0.3cm}
\end{table}
This criterion highlights the model with the highest log-likelihood.\footnote{As there are the same number of observations and the same number of parameters to estimate in the two models the BIC, for example, would have given the same results.}  We note a clear advantage for Extended Gini, dominating utilitarianism for around 70\% of participants. Once again, this is an argument in favour of the Gini index over, for example, the Atkinson index. 

\subsection{Non-parametric estimations of the extended Gini model}
\label{subsec-non_param_estim}

One of the main objectives of this paper is to see whether alternative principles to the Pigou-Dalton principle of transfers are more in line with individuals' preferences. Utilitarianism is not flexible enough, in the sense that a transfer (uniform or not) increases social welfare if and only if the utility function $u \in \ensU$ is concave. The Extended Gini model makes it possible to distinguish between the two. \cite{CM06} have established that the class of weighting functions $f \in \ensF$ consistent with URL transfers is as follows:
\begin{equation} \label{scw-URL}
\ensF_{URL} \defn \{ f \in \ensF \mid \forall t\,, f(t) \leq t \}\,.
\end{equation}
In that case, the function $f$ needs to be below the first diagonal. For the other uniform transfers, we first need to introduce the following definition. A function $f \in \ensF$ is said \textit{star-shaped} (from above) at $\xi \in [0,1]$ if and only if:
\begin{equation} \label{star-shaped}
\forall s,t \in [0,1] \setminus \xi\,,\quad s < t\ \Longrightarrow\ \frac{f(s) - f(\xi)}{s-\xi} \leq \frac{f(t) - f(\xi)}{t-\xi}\,.
\end{equation}
One obtains the following classes of transfers, respectively, for the UR and UL transfers. $\ensF_{UR}$ is the class of functions $f$ star-shaped at 0, and $\ensF_{UL}$ is the class of functions $f$ star-shaped at 1. Finally, the class of functions $f$ consistent with the Pigou-Dalton principle of transfers is denoted~$\ensF_{PT}$, and corresponds to the class of convex functions. 

In this section we propose a non-parametric (point-by-point) estimation of each participant's $f$ function, to see if one class represents their preferences better than the others. We first note that $\ensF_{PT} \subset \ensF_{UR} \subset \ensF_{URL}$ and $\ensF_{PT} \subset \ensF_{UL} \subset \ensF_{URL}$. At first sight, it may seem strange that a class of functions compatible with transfers that are more restrictive than usual PT transfers should be larger. However, this is logical given the HLP theorem (see Section~\ref{sec-framework}). In Statement (b2) of this theorem, we can associate a particular convex function with a particular social decision maker. If $\x$ is obtained from $\y$ by means of a sequence of Pigou-Dalton transfers, then Statement (b1) tells us that the social welfare function must be higher in $\x$ than in $\y$ for all social decision makers with a convex function $f$. If transfers are more restrictive in Statement (a), then it seems obvious that unanimity of rankings between $\x$ and $\y$ must be sought in a larger set of social decision makers (those accepting PT transfers, as restricted transfers are also PT transfers, plus those accepting only restricted transfers). Formally, the preorder relation identified is `less complete'.

The estimation procedure is comparable to the parametric approach. Based on Equation~\ref{param-f-full}, we have $\Delta_{f} (\x,\y; \alpha) = \alpha \sum_{i=2}^{5} \left[ f(\frac{5-i+1}{5}) - (\frac{5-i+1}{5}) \right] d_i$. By letting $\beta_i =  f(\frac{5-i+1}{5}) - (\frac{5-i+1}{5})$, we have to estimate the following (linear) model:

\begin{equation} \label{non-param-f} 
\Delta_{{f}} (\x,\y; \alpha) = \alpha \sum_{i=2}^{5} \beta_i d_i\,,
\end{equation}
where the parameters to be estimated are $\alpha > 0$ and all the $\beta_i$. The restrictions to be placed on the $\beta_i$ only guarantee that $f \in \ensF$. We deduce from these estimates the following points: $f(0.2)$, $f(0.4)$, $f(0.6)$, $f(0.8)$ recalling that, by definition, $f(0)=0$ and $f(1)=1$.

We estimate this model for each participant. As in the parametric approach, the convergence rate of the SANN algorithm is better. It is 100\%, as compared to 89.29\% for BFGS. 
We present in Table~\ref{Resultats-f-classes} the distribution of each participant, in each class of $f$ functions. 
\begin{table}[!htp] \caption{\textit{Percentage of participants in each class of weighting functions}} \label{Resultats-f-classes}
    \vspace{0.2cm}
    \centering
    \footnotesize
    \begin{tabular}{l *{4}{c}}
        \toprule
      	\textsc{Algorithm} & $f \in \ensF_{URL}$ & $f \in \ensF_{UL}$    & $f \in \ensF_{UR}$  & $f \in \ensF_{PT}$  \\
       \cmidrule(lr){1-1} \cmidrule(lr){2-5}                                                   
        SANN   & \textbf{76.34$\%$} & 57.14$\%$ & 23.44$\%$ & 10.49$\%$ \\
        BFGS   & \textbf{77.01$\%$} & 64.06$\%$ & 55.13$\%$ & 36.38$\%$ \\
        \bottomrule
    \end{tabular}\\
    \vspace{0.3cm}
\end{table}
Unsurprisingly, class~$\ensF_{URL}$ gathers the majority of participants (about 75\%). This percentage decreases with uniform transfers on one side only, with a clear advantage for uniform transfers on the left (UL). In contrast, a very large number of subjects are lost by the usual Pigou-Dalton approach: with the SANN algorithm, only 10\% of subjects are in class $\ensF_{PT}$. As a consequence, a convex function $f$ is not representative of the preferences, if each participant is considered individually. 

In order to have results comparable to those in Section~\ref{sec-acceptance}, we present in Table~\ref{Resultats-f-classes-2} the gain in number of participants that we obtain when we move from one class to a larger class, starting with $\ensF_{PT}$.
\begin{table}[!htp] \caption{\textit{Percentage of participants in each class of weighting functions}} \label{Resultats-f-classes-2}
    \vspace{0.2cm}
    \centering
    \footnotesize
    \begin{tabular}{l *{5}{c}}
        \toprule
      	\textsc{Algorithm} & $(\ensF_{URL} - \ensF_{UL})$ & $(\ensF_{UL} - \ensF_{PT})$  & $(\ensF_{URL} - \ensF_{UR})$ & $(\ensF_{UR} - \ensF_{PT})$ &  $\ensF_{PT}$ \\
       \cmidrule(lr){1-1} \cmidrule(lr){2-3} \cmidrule(lr){4-5}  \cmidrule(lr){6-6}                                                   
        SANN   & 19.20$\%$ & \textbf{46.65$\%$} & \textbf{52.90$\%$} & 12.95$\%$ & 10.49$\%$ \\
        BFGS   & 12.95$\%$ & 27.68$\%$ & 21.88$\%$ & 18.75$\%$ & 36.38$\%$ \\
        \bottomrule
    \end{tabular}\\
    \vspace{0.3cm}
\end{table}
The marginal gain from switching from $\ensF_{UL}$  to $\ensF_{URL}$  is relatively small. On the other hand, the gain between $\ensF_{UL}$  and $\ensF_{PT}$  is high. The situation is reversed if $\ensF_{UL}$  is replaced by $\ensF_{UR}$. This result reinforces our descriptive results in Section~\ref{sec-acceptance}, given an advantage to uniformity-on-the-left. 

\begin{figure}[!htp] \caption{\label{Non-param-estim} \textit{Point-by-point representation of the median weighting function}} 
\vspace{0.2cm}
\small
\begin{tabular}{cc}
\includegraphics[width=8.5cm]{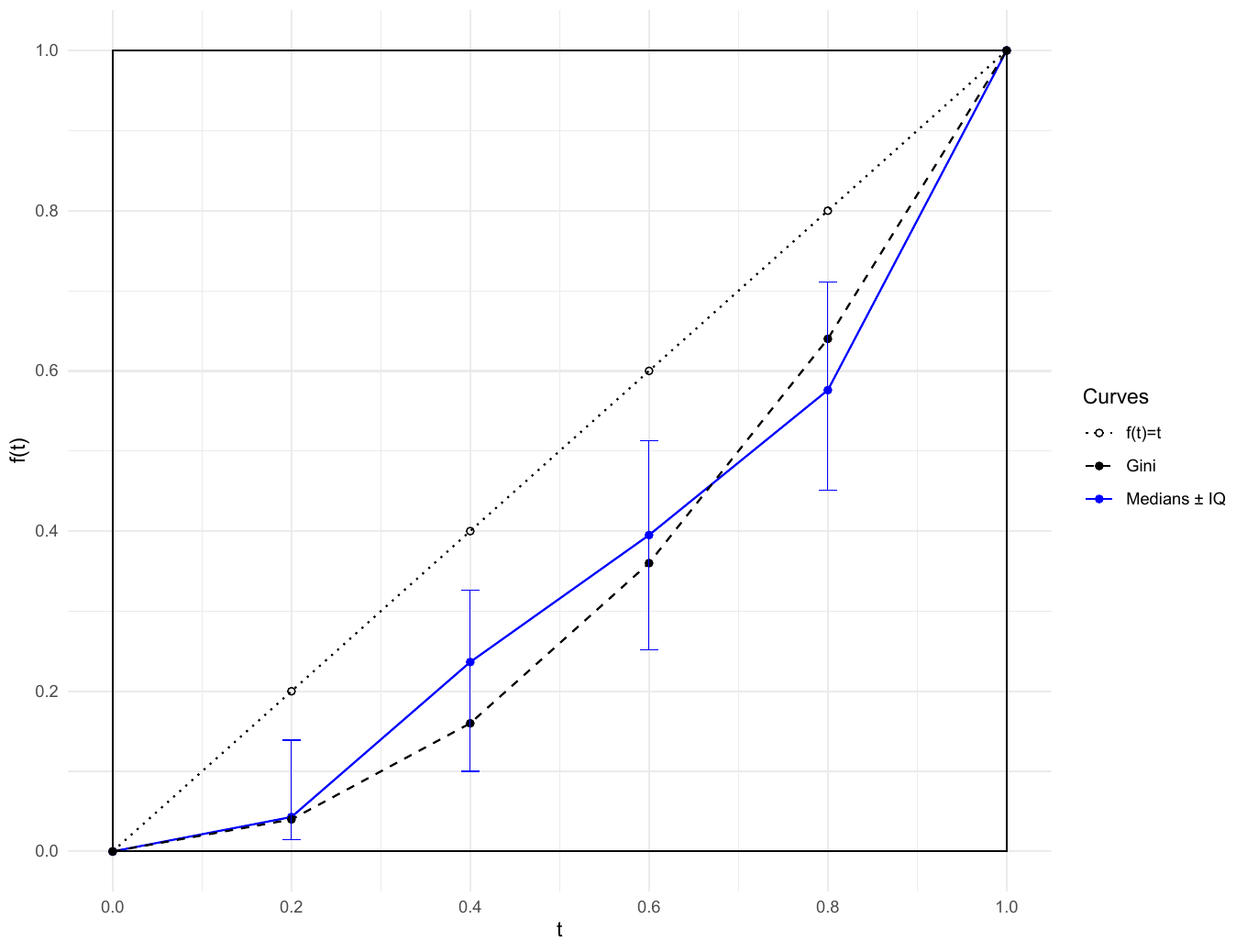} & %
\includegraphics[width=8.5cm]{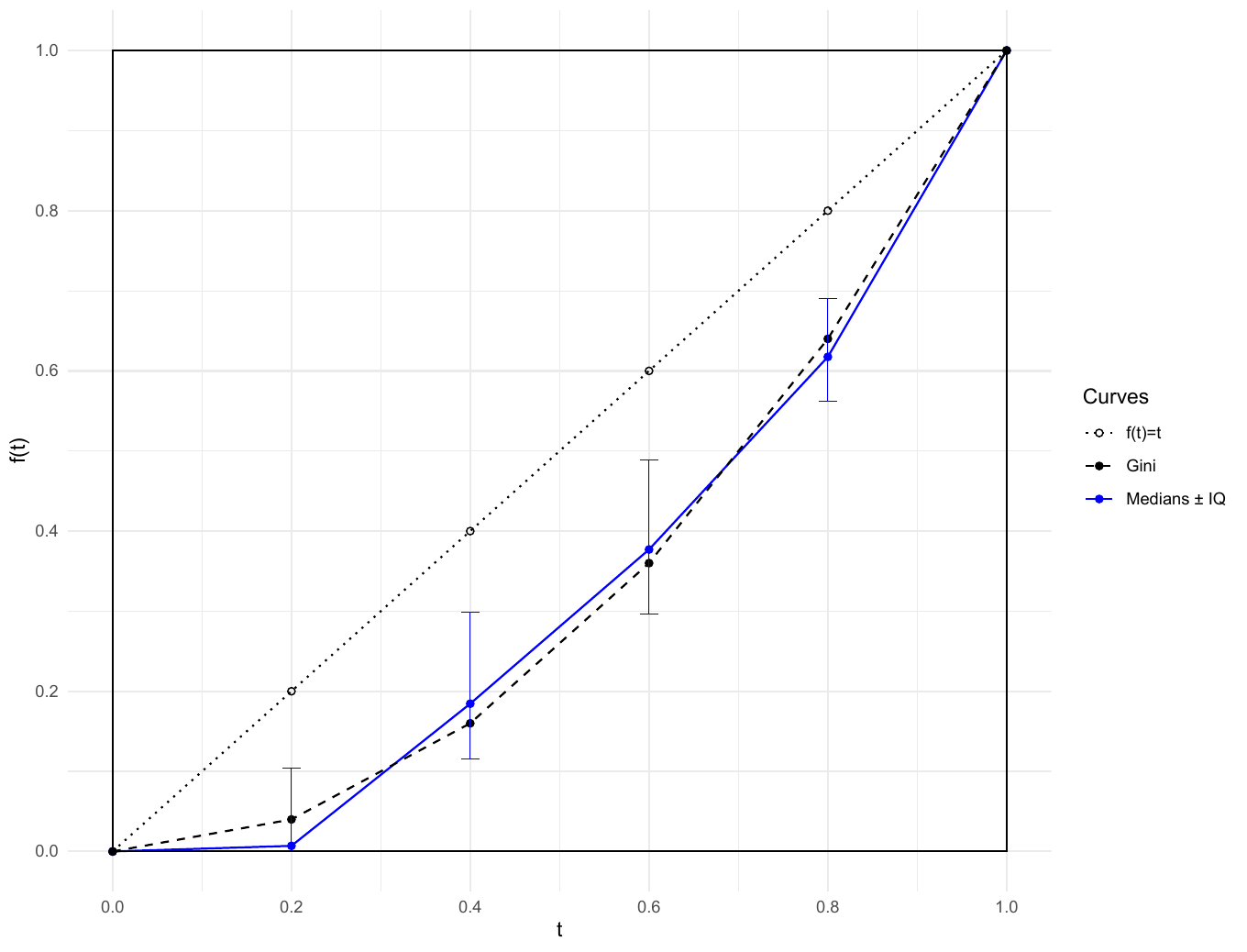} \\ 
\textit{SANN Algorithm} & \textit{BFGS Algorithm} 
\end{tabular}
\end{figure}

The values for $f(0.2)$, $f(0.4)$, $f(0.6)$, $f(0.8)$, representative of our restricted sample of participants is presented in Table~\ref{Resultats-pointbypoint}. Again we show the mean and standard-deviation, but also the median, the $1^{st}$ and $3^{rd}$ quartiles. 
\begin{table}[!htp] \caption{\textit{Representative weighting functions over all the participants}} \label{Resultats-pointbypoint}
    \vspace{0.2cm}
    \centering
    \footnotesize
    \begin{tabular}{l *{8}{c}}
        \toprule
         & \multicolumn{4}{c}{SANN \textsc{Algorithm}} & \multicolumn{4}{c}{BFGS \textsc{Algorithm}} \\
        \cmidrule(lr){2-5}  \cmidrule(lr){6-9}    
        \textsc{Statistics}   & $f(0.2)$  & $f(0.4)$ & $f(0.6)$ & $f(0.8)$ & $f(0.2)$  & $f(0.4)$ & $f(0.6)$ & $f(0.8)$   \\
        \cmidrule(lr){1-1} \cmidrule(lr){2-5}  \cmidrule(lr){6-9}                                                                                        
        Mean                          & 0.09 & 0.25 & 0.41 & 0.60 & 0.08 & 0.24 & 0.42 & 0.63 \\
        Standard-Deviation    & 0.11 & 0.19 & 0.20 & 0.18 & 0.12 & 0.18 & 0.19 & 0.17 \\
        \cmidrule(lr){1-1} \cmidrule(lr){2-5}  \cmidrule(lr){6-9}
        \textbf{Median}          & \textbf{0.04} & \textbf{0.24} & \textbf{0.40} & \textbf{0.58} & \textbf{0.01} & \textbf{0.18} & \textbf{0.38} & \textbf{0.62} \\
        1st Quartile                & 0.01 & 0.10 & 0.25 & 0.45 & 0.00 & 0.12 & 0.30 & 0.56 \\
        3rd Quartile                & 0.14 & 0.33 & 0.51 & 0.71 & 0.10 & 0.30 & 0.49 & 0.69 \\      
        \cmidrule(lr){1-1} \cmidrule(lr){2-5}  \cmidrule(lr){6-9}
        \textbf{Gini - Median} & \textbf{0} & \textbf{-0.08} & \textbf{-0.04} & \textbf{0.06} & \textbf{0.03} & \textbf{-0.02} & \textbf{-0.02} & \textbf{0.02} \\
        \bottomrule
    \end{tabular}\\
    \vspace{0.3cm}
\end{table}
We also compute the difference with the different values for the Gini index, recalling that it is characterised by $f(t)=t^2$. The SANN and BFGS algorithms provide comparable results. Moreover, and surprisingly enough, the median values of each computed point for $f(t)$ are almost identical to those of the Gini index. This is surprising because convex weighting functions $f$ poorly represent the preferences of each participant, taken individually. We plot the estimates in Figure~\ref{Non-param-estim}, including the \textit{Interquartile Range} (Q3-Q1) for each point $f(t)$. 
The result observed in the parametric approach is repeated here. The $f$ function characterising the Gini index very faithfully represents the preferences of the median individual (particularly with the BFGS algorithm). However, we must qualify this observation slightly, particularly for the estimate produced by the SANN algorithm. We check in Table~\ref{Resultats-f-classes-median} to which class the non-parametric function $f$ of the median individual belongs.
\begin{table}[!htp] \caption{\textit{Median individual preferences and weighting functions classes}} \label{Resultats-f-classes-median}
    \vspace{0.2cm}
    \centering
    \footnotesize
    \begin{tabular}{l *{4}{c}}
        \toprule
      	\textsc{Algorithm} & $f \in \ensF_{URL}$ & $f \in \ensF_{UL}$    & $f \in \ensF_{UR}$  & $f \in \ensF_{PT}$  \\
       \cmidrule(lr){1-1} \cmidrule(lr){2-5}                                                   
        SANN   & Yes & Yes & Yes & \textbf{No} \\
        BFGS   & Yes & Yes & Yes & Yes \\
        \bottomrule
    \end{tabular}\\
    \vspace{0.3cm}
\end{table}
If it is indeed convex for BFGS, it is star-shaped at 0 and 1 for SANN (i.e. in classes $\ensF_{URL}$, $\ensF_{UL}$ and $\ensF_{UR}$), but not convex.

\section{Discussion}
\label{sec-discussion}

The results of our web-experiment indicate a clear rejection of progressive transfers in the middle of the income distribution--those that do not involve the poorest or the richest individuals. A majority of participants in our study believe that such transfers do not reduce overall inequality. This finding aligns with trends observed in other experimental analyses, although previous studies did not formally distinguish between uniform and non-uniform transfers. In contrast, transfers that promote solidarity at both the top and bottom of the distribution (referred to as URL transfers) receive significantly higher approval in our sample. Additionally, we observe a stronger preference for solidarity among the poor (UL transfers) compared to that among the rich (UR transfers), indicating a greater prioritization of poverty reduction over the mitigation of extreme wealth. This also suggests that the theoretical distinction between uniform and non-uniform transfers is relevant in the context of our experiment.

The other question investigated in this paper, related to the previous observations, concerns the relevance of the inequality indices used in the empirical literature, in light of individual preferences. Since all standard indices conform to the Pigou-Dalton principle of transfers, which is rejected by a majority of our subjects, their legitimacy is called into question. First, the Extended Gini Donaldson-Weymark approach fits individual preferences much better than indices of the Atkinson-Kolm-Sen class. Focusing on the first approach, we observe that the subclass compatible with the principle of transfers (characterized by a convex weighting function) poorly aligns with preferences when considering individuals in isolation. As a consequence, the Gini index represents the preferences of very few individuals. However, when we focus on the median individual preferences, the Gini index adjusts relatively well. These results are summarised in Figure~\ref{Param-estim} for parametric estimates, and in Figure~\ref{Non-param-estim} for non-parametric estimates.
 
 Based on these results, several research avenues can be explored. First, the higher acceptance of transfer principles that impose solidarity among donors or recipients, warrants confirmation through various experimental designs and diverse subject populations. Another potential project could involve identifying and empirically estimating a parametric class of star-shaped weighting functions $f \in \ensF$ that better align with individual preferences. Such a model could also more accurately capture the preferences of the median individual, as estimated, for instance, by the SANN algorithm in Figure~\ref{Non-param-estim}. 

\bibliographystyle{ecca}
\bibliography{Biblio_202411}

\newpage
\appendix
\section{Appendix}

\subsection{Instructions provided to respondents (translated from French)}
\label{ap-instructions}

\footnotesize

\noindent \textbf{SCREEN 1: General description of the study}.

\noindent The study you have agreed to take part in is being carried out by several French university research centres, specialised in the study of inequality. Our aim is to construct inequality indicators in order to measure the impact on income distribution of government interventions in the economic sphere. These interventions may concern areas as diverse as taxation, family policy, the pension system, housing subsidies and the financing of the healthcare system, to name but a few. 

\noindent These different government interventions are likely to modify the incomes of members of society. We believe that the indicators used to assess the impact of these interventions on income distribution should reflect, as far as possible, the point of view of members of society, who are the first to be affected. Your participation in this study will enable us to gather a range of opinions representative of the different points of view on inequality within French society. 

\vspace{0.2cm}
\noindent \textbf{SCREEN 2: This study is in three parts}.

\noindent The first part consists of 44 questions. For each question, we will present you with two income distributions and ask you to indicate which of these two distributions you think is the least unequal. 

\noindent After each group of 11 questions, you will be able to consult your answers and, if you wish, modify them. We would like to stress that there are no right or wrong answers: we are only interested in your personal opinion. 

\noindent In the second part, we will ask you whether you agree or disagree with a number of statements about the impact on inequality of different types of income redistribution between individuals. Again, there is no right or wrong answer: you are free to agree or disagree with the statements.  

\noindent In the third part, we will ask you a series of personal questions to help us situate you in French society. The aim here is to ensure that all the people who took part in this study are as faithful a representation of French society as possible. 

\noindent We would like to stress that your answers will remain anonymous. Similarly, all personal information collected will remain confidential. It will only be used for our research work and it will not be possible to identify you from the information collected. It is imperative for the success of this study that you take the utmost care when reading the questions and answering them. It is also important that you complete the questionnaire to the end. We estimate that the average time spent answering the questionnaire should not exceed 30 minutes. When we have completed our survey, you will receive an e-mail with a link to the results. 
	
\vspace{0.2cm}
\noindent \textbf{SCREEN 3: Part One}.

\noindent Imagine a society consisting of 5 perfectly identical individuals: there are no personal characteristics to distinguish them from one another. There is no reason why they should be treated differently. 

\noindent We are interested in the level of inequality in this society by considering only the income of individuals, expressed in thousands of euros. In each question in this first part, two competing economic policies are considered, each leading to a particular income distribution: Distribution A and Distribution B. 

\noindent The sum of distributed income is the same in both distributions. 

\noindent You are asked to compare these two distributions from the point of view of inequality:
\begin{itemize}
\item[-]  If you consider that Distribution A is less unequal than Distribution B, then tick the `Distribution A' box. 
\item[-] If you consider that Distribution B is less unequal than Distribution A, then tick the `Distribution B' box.  
\item[-] Finally, if you are unable to decide, or if you consider that the two distributions are equivalent, then tick the `Equivalent' box. 
\end{itemize}

\noindent Sample question:
\begin{center}
In your opinion, which distribution is the least unequal?	
\end{center}

\begin{center}
\begin{tabular}{*{3}{c}}
        \toprule
Distribution A & \multirow{2}{*}{Equivalent}  & Distribution B \\
(2,6,10,14,18) & & (3,6,10,14,17) \\
\bottomrule
\end{tabular}
\end{center}

\noindent Reading: Distribution A gives an income of \euro2,000 to the 1st person,  \euro6,000 to the 2nd person,  \euro10,000 to the 3rd person,  \euro14,000 to the 4th person and  \euro18,000 to the 5th person.

\noindent The questionnaire will now begin (then, presentation of the 44 questions, one screen for each question). 

\vspace{0.2cm}
\noindent \textbf{SCREEN 4: Part Two}.

\noindent Here again we are considering a fictive society consisting of perfectly identical individuals: there is still no reason to favour one individual over another.  You are asked to indicate the extent to which you agree with a number of statements concerning the impact on inequality of different ways of redistributing income between individuals.

\noindent \# To the following statement: `a transfer of income from individual X to individual Y (who is poorer than the first) always reduces inequality in society as a whole', do you...?

\vspace{-0.4cm}
\begin{center}
Strongly disagree / Somewhat disagree / No opinion / Somewhat agree / Strongly agree
\end{center}

\noindent \# To the following statement: `a transfer of income from individual X to individual Y (poorer than the former) reduces inequality in society as a whole, on the sole condition that individuals poorer than Y receive at least the same amount of income as that received by Y', do you :

\vspace{-0.4cm}
\begin{center}
Strongly disagree / Somewhat disagree / No opinion / Somewhat agree / Strongly agree
\end{center}

\noindent \# To the following statement: `a transfer of income from an individual X to an individual Y (poorer than the former) reduces inequalities in society as a whole, on the sole condition that individuals richer than X give at least the same amount of income as that given by X', do you :

\vspace{-0.4cm}
\begin{center}
Strongly disagree / Somewhat disagree / No opinion / Somewhat agree / Strongly agree
\end{center}

\noindent \# To the following statement: `a transfer of income from an individual X to an individual Y (poorer than the former) reduces inequality in society as a whole, on the sole conditions that (a) individuals poorer than Y receive at least the same amount of income as that received by Y and (b) individuals richer than X give at least the same amount of income as that given by X', do you :

\vspace{-0.4cm}
\begin{center}
Strongly disagree / Somewhat disagree / No opinion / Somewhat agree
\end{center}

\newpage

\subsection{List of the numerical questions}
\label{liste_question}
%
\begin{table}[!htp] \caption{List of the 55 possible questions} \label{list_question-tab}
    \vspace{0.2cm}
    \centering
    \scriptsize
    \begin{tabular}{cl*{15}{c}}
        \toprule
       \textsc{Initial distrib.} & \textsc{Final distrib.} & \multicolumn{5}{c}{\textsc{Distribution A}} & \multicolumn{5}{c}{\textsc{Distribution B}} &
        \multicolumn{5}{c}{\textsc{Transfer}} \\
        \cmidrule(lr){1-1}  \cmidrule(lr){2-2} \cmidrule(lr){3-7}	\cmidrule(lr){8-12} \cmidrule(lr){13-17}
        \multirow{10}{*}{$\y^1$}
       & $\y^1 +\xt^1$                 & 2 & 6   & 10 & 14 & 18                       & 3 & 6   & 10 & 14 & 17		& +1 &  0  &  0  &  0  & -1 \\
       
       & $\y^1 +\xt^2$                 & 2 & 6   & 10 & 14 & 18                       & 2 & 7   & 10 & 14 & 17		&  0  & +1 &  0  &  0  & -1 \\
       & $\y^1 +\xt^3$                 & 2 & 6   & 10 & 14 & 18                       & 2 & 6   & 11 & 14 & 17		&  0  &  0  & +1 &  0  & -1 \\
       & $\y^1 +\xt^4$                 & 2 & 6   & 10 & 14 & 18                       & 2 & 6   & 10 & 15 & 17		&  0  &  0  &  0  & +1 & -1 \\
       
       & $\y^1 +\xt^5$                 & 2 & 6   & 10 & 14 & 18                       & 3 & 6   & 10 & 13 & 18		& +1 &  0  &  0  & -1 &  0  \\
       & $\y^1 +\xt^6$                 & 2 & 6   & 10 & 14 & 18                       & 3 & 6   & 9   & 14 & 18		& +1 &  0  & -1  &  0 &  0  \\
       & $\y^1 +\xt^7$                 & 2 & 6   & 10 & 14 & 18                       & 3 & 5   & 10 & 14 & 18		& +1 & -1  &  0  &  0 &  0  \\
       
       & $\y^1 +\xt^8$                 & 2 & 6   & 10 & 14 & 18                       & 2 & 6   & 11 & 13 & 18		&  0  &  0  & +1 & -1  & 0  \\
       & $\y^1 +\xt^9$                 & 2 & 6   & 10 & 14 & 18                       & 2 & 7   & 9   & 14 & 18		&  0  & +1 & -1  &  0  & 0  \\
       & $\y^1 +\xt^{10}$               & 2 & 6   & 10 & 14 & 18                       & 2 & 7   & 10 & 13 & 18		&  0  & +1 &  0  & -1  & 0  \\
       
       & TEST $\y^1$             & 2 & 6   & 10 & 14 & 18                       & 10 & 10 & 10 & 10 & 10 \\
       
       \cmidrule(lr){1-1}  \cmidrule(lr){2-2} \cmidrule(lr){3-7}	\cmidrule(lr){8-12} \cmidrule(lr){13-17}
       \multirow{10}{*}{$\y^2$}
       & $\y^2 +\xt^1$                 & 2 & 4   & 14 & 16 & 18                        & 3 & 4   & 14 & 16 & 17		& +1 &  0  &  0  &  0  & -1 \\
       
       & $\y^2 +\xt^2$                 & 2 & 4   & 14 & 16 & 18                        & 2 & 5   & 14 & 16 & 17		&  0  & +1 &  0  &  0  & -1 \\
       & $\y^2 +\xt^3$                 & 2 & 4   & 14 & 16 & 18                        & 2 & 4   & 15 & 16 & 17		&  0  &  0  & +1 &  0  & -1 \\
       & $\y^2 +\xt^4$                 & 2 & 4   & 14 & 16 & 18                        & 2 & 4   & 14 & 17 & 17		&  0  &  0  &  0  & +1 & -1 \\
       
       & $\y^2 +\xt^5$                 & 2 & 4   & 14 & 16 & 18                        & 3 & 4   & 14 & 15 & 18		& +1 &  0  &  0  & -1 &  0  \\
       & $\y^2 +\xt^6$                 & 2 & 4   & 14 & 16 & 18                        & 3 & 4   & 13 & 16 & 18		& +1 &  0  & -1  &  0 &  0  \\
       & $\y^2 +\xt^7$                 & 2 & 4   & 14 & 16 & 18                        & 3 & 3   & 14 & 16 & 18		& +1 & -1  &  0  &  0 &  0  \\
       
       & $\y^2 +\xt^8$                 & 2 & 4   & 14 & 16 & 18                        & 2 & 4   & 15 & 15 & 18		&  0  &  0  & +1 & -1  & 0  \\
       & $\y^2 +\xt^9$                 & 2 & 4   & 14 & 16 & 18                        & 2 & 5   & 13 & 16 & 18		&  0  & +1 & -1  &  0  & 0  \\
       & $\y^2 +\xt^{10}$               & 2 & 4   & 14 & 16 & 18                        & 2 & 5   & 14 & 15 & 18		&  0  & +1 &  0  & -1  & 0  \\
       
       & TEST $\y^2$             & 2 & 4   & 14 & 16 & 18                       & 10 & 10 & 10 & 10 & 10 \\
              
       \cmidrule(lr){1-1}  \cmidrule(lr){2-2} \cmidrule(lr){3-7}	\cmidrule(lr){8-12} \cmidrule(lr){13-17}
       \multirow{10}{*}{$\y^3$}
       & $\y^3 +\xt^1$                 & 2 & 4   & 6   & 16 & 18                        & 3 & 4   & 6   & 16 & 17		& +1 &  0  &  0  &  0  & -1 \\
       
       & $\y^3 +\xt^2$                 & 2 & 4   & 6   & 16 & 18                        & 2 & 5   & 6   & 16 & 17		&  0  & +1 &  0  &  0  & -1 \\
       & $\y^3 +\xt^3$                 & 2 & 4   & 6   & 16 & 18                        & 2 & 4   & 7   & 16 & 17		&  0  &  0  & +1 &  0  & -1 \\
       & $\y^3 +\xt^4$                 & 2 & 4   & 6   & 16 & 18                        & 2 & 4   & 6   & 17 & 17		&  0  &  0  &  0  & +1 & -1 \\
       
       & $\y^3 +\xt^5$                 & 2 & 4   & 6   & 16 & 18                        & 3 & 4   & 6   & 15 & 18		& +1 &  0  &  0  & -1 &  0   \\
       & $\y^3 +\xt^6$                 & 2 & 4   & 6   & 16 & 18                        & 3 & 4   & 5   & 16 & 18		& +1 &  0  & -1  &  0 &  0   \\
       & $\y^3 +\xt^7$                 & 2 & 4   & 6   & 16 & 18                        & 3 & 3   & 6   & 16 & 18		& +1 & -1  &  0  &  0 &  0   \\
       
       & $\y^3 +\xt^8$                 & 2 & 4   & 6   & 16 & 18                        & 2 & 4   & 7   & 15 & 18		&  0  &  0  & +1 & -1  & 0  \\
       & $\y^3 +\xt^9$                 & 2 & 4   & 6   & 16 & 18                        & 2 & 5   & 5   & 16 & 18		&  0  & +1 & -1  &  0  & 0  \\
       & $\y^3 +\xt^{10}$               & 2 & 4   & 6   & 16 & 18                        & 2 & 5   & 6   & 15 & 18		&  0  & +1 &  0  & -1  & 0  \\
       
       & TEST $\y^3$            & 2 & 4   & 6   & 16 & 18                       & 10 & 10 & 10 & 10 & 10 \\
       
       \cmidrule(lr){1-1}  \cmidrule(lr){2-2} \cmidrule(lr){3-7}	\cmidrule(lr){8-12} \cmidrule(lr){13-17}
       \multirow{10}{*}{$\y^4$}
       & $\y^4 +\xt^1$                 & 2 & 8   & 10 & 12 & 18                        & 3 & 8   & 10 & 12 & 17		& +1 &  0 &  0  &  0  & -1 \\
       
       & $\y^4 +\xt^2$                 & 2 & 8   & 10 & 12 & 18                        & 2 & 9   & 10 & 12 & 17		&  0  & +1 &  0  &  0  & -1 \\
       & $\y^4 +\xt^3$                 & 2 & 8   & 10 & 12 & 18                        & 2 & 8   & 11 & 12 & 17		&  0  &  0  & +1 &  0  & -1 \\
       & $\y^4 +\xt^4$                 & 2 & 8   & 10 & 12 & 18                        & 2 & 8   & 10 & 13 & 17		&  0  &  0  &  0  & +1 & -1 \\
       
       & $\y^4 +\xt^5$                 & 2 & 8   & 10 & 12 & 18                        & 3 & 8   & 10 & 11 & 18		& +1 &  0  &  0  & -1 &  0   \\
       & $\y^4 +\xt^6$                 & 2 & 8   & 10 & 12 & 18                        & 3 & 8   & 9   & 12 & 18		& +1 &  0  & -1  &  0 &  0   \\
       & $\y^4 +\xt^7$                 & 2 & 8   & 10 & 12 & 18                        & 3 & 7   & 10 & 12 & 18		& +1 & -1  &  0  &  0 &  0   \\
       
       & $\y^4 +\xt^8$                 & 2 & 8   & 10 & 12 & 18                        & 2 & 8   & 11 & 11 & 18		&  0  &  0  & +1 & -1  & 0  \\
       & $\y^4 +\xt^9$                 & 2 & 8   & 10 & 12 & 18                        & 2 & 9   & 9   & 12 & 18		&  0  & +1 & -1  &  0  & 0  \\
       & $\y^4 +\xt^{10}$               & 2 & 8   & 10 & 12 & 18                        & 2 & 9   & 10 & 11 & 18		&  0  & +1 &  0  & -1  & 0  \\ 
       
       & TEST $\y^4$             & 2 & 8   & 10 & 12 & 18                       & 10 & 10 & 10 & 10 & 10 \\
            
      \cmidrule(lr){1-1}  \cmidrule(lr){2-2} \cmidrule(lr){3-7}	\cmidrule(lr){8-12} \cmidrule(lr){13-17}
       \multirow{10}{*}{$\y^5$}
       & $\y^5 +\xt^1$                 & 2 & 4   & 10 & 16 & 18                        & 3 & 4   & 10 & 16 & 17		& +1 &  0  &  0  &  0 & -1 \\
       
       & $\y^5 +\xt^2$                 & 2 & 4   & 10 & 16 & 18                        & 2 & 5   & 10 & 16 & 17		&  0  & +1 &  0  &  0  & -1 \\
       & $\y^5 +\xt^3$                 & 2 & 4   & 10 & 16 & 18                        & 2 & 4   & 11 & 16 & 17		&  0  &  0  & +1 &  0  & -1 \\
       & $\y^5 +\xt^4$                 & 2 & 4   & 10 & 16 & 18                        & 2 & 4   & 10 & 17 & 17		&  0  &  0  &  0  & +1 & -1 \\
       
       & $\y^5 +\xt^5$                 & 2 & 4   & 10 & 16 & 18                        & 3 & 4   & 10 & 15 & 18		& +1 &  0  &  0  & -1 &  0   \\
       & $\y^5 +\xt^6$                 & 2 & 4   & 10 & 16 & 18                        & 3 & 4   & 9   & 16 & 18		& +1 &  0  & -1  &  0 &  0   \\
       & $\y^5 +\xt^7$                 & 2 & 4   & 10 & 16 & 18                        & 3 & 3   & 10 & 16 & 18		& +1 & -1  &  0  &  0 &  0   \\
       
       & $\y^5 +\xt^8$                 & 2 & 4   & 10 & 16 & 18                        & 2 & 4   & 11 & 15 & 18		&  0  &  0  & +1 & -1  & 0  \\
       & $\y^5 +\xt^9$                 & 2 & 4   & 10 & 16 & 18                        & 2 & 5   & 9   & 16 & 18		&  0  & +1 & -1  &  0  & 0  \\
       & $\y^5 +\xt^{10}$               & 2 & 4   & 10 & 16 & 18                        & 2 & 5   & 10 & 15 & 18		&  0  & +1 &  0  & -1  & 0  \\
       
       & TEST $\y^5$           & 2 & 6   & 10 & 14 & 18                       & 10 & 10 & 10 & 10 & 10 \\
       
        \bottomrule
	\end{tabular}
\end{table}

\newpage

\subsection{Details on the sample of participants}
\label{ap-survey}

\begin{table}[!htp] 
\caption{Sociodemographic variables and descriptive statistics (15 years and over)} 
\label{socio-demo-tab} 
\vspace{0.2cm} 
\centering 
\scriptsize 
\begin{tabular}{*{8}{l}} 
\toprule 
\multicolumn{2}{c}{\textsc{Variables}} & \multicolumn{2}{c}{\textsc{Full sample}} & \multicolumn{2}{c}{\textsc{Restricted}} & \textsc{(R-F)} & \textsc{INSEE}\\ 
\cmidrule(lr){1-2} \cmidrule(lr){3-4} \cmidrule(lr){5-6} \cmidrule(lr){7-7} \cmidrule(lr){8-8} 
\textsc{Name} & \textsc{Value} & Nb. & \% & Nb. & \% & \% & \% \\ 
\cmidrule(lr){1-2} \cmidrule(lr){3-4} \cmidrule(lr){5-6} \cmidrule(lr){7-7} \cmidrule(lr){8-8}  
\multirow{2}{*}{Gender} & Woman & 531 & 51.65 & 232 & 51.79 & \ 0.14 & 51.6\\ 
& Man & 497 & 48.35 & 216 & 48.21 & -0.14 & 48.4 \\ 
\cmidrule(lr){1-2} \cmidrule(lr){3-4} \cmidrule(lr){5-6} \cmidrule(lr){7-7} \cmidrule(lr){8-8}  
\multirow{6}{*}{Age} & 15 -- 29 years & 123 & 11.96 & 53 & 11.83 & -0.13 & 21.2 \\ 
& 30 -- 44 years & 236 & 22.96 & 102 & 22.77 & -0.19 & 22.5 \\ 
& 45 -- 59 years & 281 & 27.33 & 105 & 23.44 & -3.89 & 23.9 \\ 
& 60 -- 74 years & 240 & 23.35 & 108 & 24.11 & \ 0.76 & 20.7 \\ 
& 75 -- 89 years & 147 & 14.30 & 79 & 17.63 & \ 3.33 & 10.0 \\ 
& $\geq$ 90 years & 1 & 00.01 & 1 & 00.22 & \ 0.21 & 01.6 \\ 
\cmidrule(lr){1-2} \cmidrule(lr){3-4} \cmidrule(lr){5-6} \cmidrule(lr){7-7} \cmidrule(lr){8-8}  
\multirow{5}{*}{Number of children} & No children & 384 & 37.35 & 172 & 38.39 & \ 1.04 & 49.4 \\ 
& 1 child & 190 & 18.48 & 76 & 16.96 & -1.52 & 21.9 \\ 
& 2 children & 285 & 27.72 & 128 & 28.57 & \ 0.85 & 19.6 \\ 
& 3 children & 125 & 12.16 & 46 & 10.27 & -1.89 & 06.7 \\ 
& 4 children or more & 44 & 04.28 & 26 & 05.81 & \ 1.53 & 02.4 \\ 
\cmidrule(lr){1-2} \cmidrule(lr){3-4} \cmidrule(lr){5-6} \cmidrule(lr){7-7} \cmidrule(lr){8-8}  
\multirow{4}{*}{Marital status} & Married/Civil-union & 516 & 50.19 & 230 & 51.34 & \ 1.15 & 47.5 \\ 
& Cohabiting/Common-law & 101 & 09.82 & 50 & 11.16 & \ 1.34 & 11.0 \\ 
& Widower & 37 & 03.60 & 18 & 04.02 & \ 0.42 & 06.8 \\ 
& Single & 374 & 36.38 & 150 & 33.48 & -2.90 & 34.7\\ 
\cmidrule(lr){1-2} \cmidrule(lr){3-4} \cmidrule(lr){5-6} \cmidrule(lr){7-7} \cmidrule(lr){8-8}  
\multirow{5}{*}{Employment status} & Employed & 530 & 51.56 & 221 & 49.34 & -2.22 & 49.3 \\ 
& Active but unemployed & 76 & 07.39 & 25 & 05.58 & -1.81 & 06.8 \\ 
& Student & 66 & 06.42 & 34 & 07.59 & \ 1.17 & 07.9  \\ 
& Retired & 278 & 27.04 & 140 & 31.25 & \ 4.21 & 29.4 \\ 
& Other inactivity situation & 78 & 07.59 & 28 & 06.25 & -1.34 & 06.6 \\ 
\cmidrule(lr){1-2} \cmidrule(lr){3-4} \cmidrule(lr){5-6} \cmidrule(lr){7-7} \cmidrule(lr){8-8}  
\multirow{7}{*}{Occupation category} & Farmers & 16 & 01.56 & 7 & 01.56 & \ 0.00 & 01.3 \\ 
& Artisans/shopkeepers/company owners & 49 & 04.77 & 23 & 05.13 & \ 0.36 & 06.6 \\ 
& Managers/higher intellectual professions & 217 & 21.11 & 124 & 27.68 & \ 6.57 & 17.2 \\ 
& Intermediate occupations & 215 & 20.91 & 105 & 23.44 & \ 2.53 & 22.4 \\ 
& Employees & 297 & 28.89 & 115 & 25.67 & -3.22 & 22.6 \\ 
& Manual workers & 145 & 14.11 & 35 & 07.81 & -6.30 & 19.9 \\ 
& Not concerned & 89 & 08.66 & 39 & 08.71 & \ 0.05 & 10.0 \\ 
\cmidrule(lr){1-2} \cmidrule(lr){3-4} \cmidrule(lr){5-6} \cmidrule(lr){7-7} \cmidrule(lr){8-8}  
\multirow{6}{*}{Education} & Primary education & 34 & 03.31 & 6 & 01.34 & -1.97 & 22.2 \\ 
& Lower secondary education & 92 & 08.95 & 18 & 04.02 & -4.93 & 06.0 \\ 
& Upper secondary education & 338 & 32.88 & 123 & 27.46 & -5.42 & 38.2  \\ 
& Short cycle tertiary education & 224 & 21.79 & 98 & 21.88 & \ 0.09 & 11.8 \\ 
& Bachelor & 144 & 14.01 & 84 & 18.75 & \ 4.74 & 11.2 \\ 
& Master/Doctorate & 196 & 19.07 & 119 & 26.56 & \ 7.49 & 10.6 \\ 
\cmidrule(lr){1-2} \cmidrule(lr){3-4} \cmidrule(lr){5-6} \cmidrule(lr){7-7} \cmidrule(lr){8-8}  
\multirow{10}{*}{Gross monthly income} & $\leq$ \euro1,200 & 132 & 12.84 & 43 & 09.60 & -3.24 & D1 \\ 
& \euro1,201 -- \euro1,500 & 113 & 10.99 & 36 & 08.04 & -2.95 & D2 \\ 
& \euro1,501 -- \euro1,800 & 85 & 08.27 & 33 & 07.37 & -0.90 & D3 \\ 
& \euro1,801 -- \euro2,200 & 112 & 10.89 & 56 & 12.50 & \ 1.61 & D4 \\ 
& \euro2,201 -- \euro2,600 & 117 & 11.38 & 52 & 11.61 & \ 0.23 & D5 \\ 
& \euro2,601 -- \euro3,000 & 104 & 10.12 & 38 & 08.48 & -1.64 & D6 \\ 
& \euro3,001 -- \euro3,500 & 90 & 08.75 & 34 & 07.59 & -1.16 & D7 \\ 
& \euro3,501 -- \euro4,200 & 118 & 11.48 & 67 & 14.96 & \ 3.48 & D8 \\ 
& \euro4,201 -- \euro5,400 & 93 & 09.05 & 58 & 12.95 & \ 3.90 & D9 \\ 
& $>$ \euro5,400 & 64 & 06.23 & 31 & 06.92 & \ 0.69 & D10 \\ 
\cmidrule(lr){1-2} \cmidrule(lr){3-4} \cmidrule(lr){5-6} \cmidrule(lr){7-7} \cmidrule(lr){8-8}  
\multirow{2}{*}{Vote last presidential election} & Yes & 847 & 82.39 & 381 & 85.04 & \ 2.65 & 85.0 \\ 
& No & 181 & 17.61 & 67 & 14.96 & -2.65 & 15.0 \\ 
\cmidrule(lr){1-2} \cmidrule(lr){3-4} \cmidrule(lr){5-6} \cmidrule(lr){7-7} \cmidrule(lr){8-8}  
\multirow{6}{*}{Political opinion} & Do not wish to reply & 338 & 32.88 & 99 & 22.10 & -10.78 & -- \\ 
& Extreme left & 21 & 02.04 & 15 & 03.35 & \ 1.31 & -- \\ 
& Left & 224 & 21.79 & 121 & 27.01 & \ 5.22 & --  \\ 
& Centre & 214 & 20.82 & 120 & 26.79 & \ 5.97 & -- \\ 
& Right & 162 & 15.76 & 73 & 16.29 & \ 0.53 & -- \\ 
& Extreme Right & 69 & 06.71 & 20 & 04.46 & -2.25 & -- \\ 
\cmidrule(lr){1-2} \cmidrule(lr){3-8} 
\multicolumn{2}{c}{Number of observations} & 1028 & & 448 & & & \\ 
\bottomrule 
\end{tabular}\\
    \vspace{0.3cm}
    \small \textit{Notes}. `Restricted sample' corresponds to the participants who have correctly answered to all the test questions. INSEE Data come from `\href{https://www.insee.fr/fr/statistiques/2011101?geo=FRANCE-1}{\textit{RP2021, exploitations principales, géographie au 01/01/2024}}'.
\end{table}

\begin{figure}[!htp] \caption{\textit{Difficulties encountered by the participants in the questions}}\label{Plot-errors}
\small
\begin{center}
\includegraphics[width=12cm]{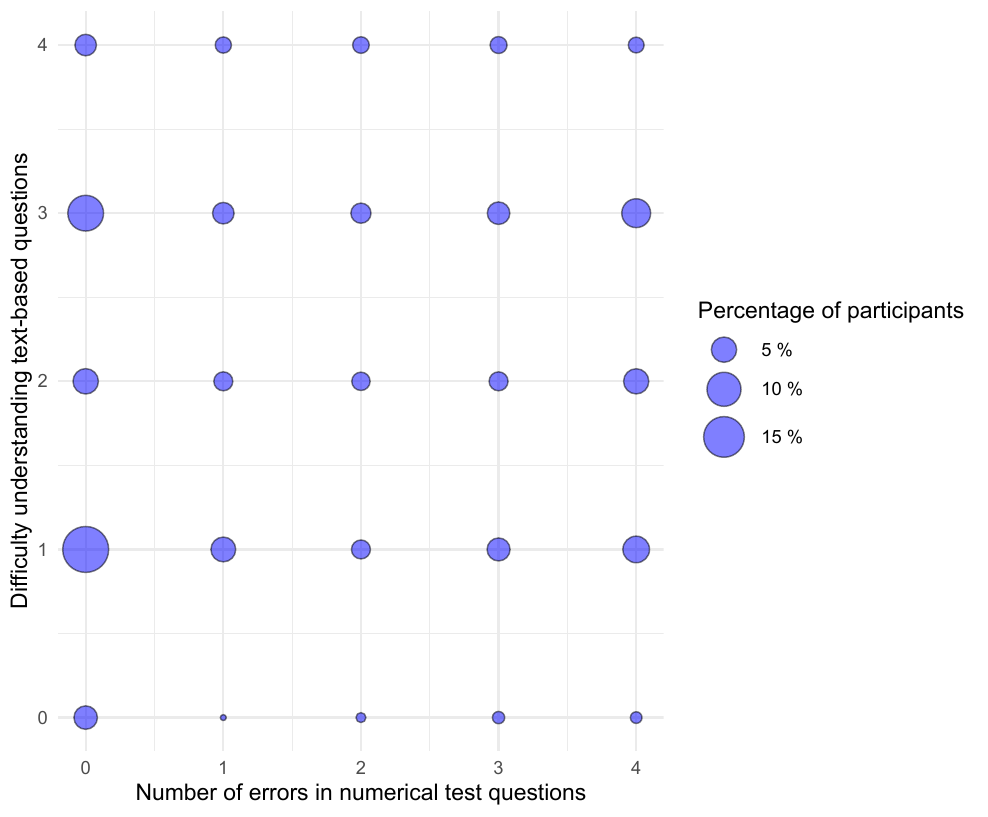} \\ 
    \vspace{0.3cm}
    \small \textit{Notes}. On the y-axis: 0 = `Really clear' to 4 =`Not clear at all'.
\end{center}

\end{figure} 

\newpage
\subsection{Acceptation rates by socio-economic characteristics}
\label{other-tables}

\begin{table}[!htp] \caption{\textit{Acceptation rates by gender}} \label{Resulats-genre}
    \vspace{0.2cm}
    \centering
    \footnotesize
\begin{tabular}{ll*{4}{c}}
\toprule
\textsc{Transfers} & & \textsc{Accepted} & \textsc{Rejected} & \textsc{Neutrality} & \textit{N} \\
	\cmidrule(lr){1-1} \cmidrule(lr){2-2} \cmidrule(lr){3-5}  \cmidrule(lr){6-6}
\multirow{2}{*}{URL} & Women &    61.31$\%$&    8.19$\%$&    30.50$\%$& 928\\
& Men & 72.69$\%$ & 8.68$\%$ & 18.63$\%$ & 864\\ 
	\cmidrule(lr){1-1} \cmidrule(lr){2-2} \cmidrule(lr){3-5}  \cmidrule(lr){6-6}
\multirow{2}{*}{UL} & Women &    56.18$\%$ &    10.09$\%$ &    33.73$\%$ & 2784\\
& Men & 62.65$\%$ & 14.08$\%$ & 23.26$\%$ & 2592\\ 
	\cmidrule(lr){1-1} \cmidrule(lr){2-2} \cmidrule(lr){3-5}  \cmidrule(lr){6-6}
\multirow{2}{*}{UR} & Women &    47.59$\%$&    12.79$\%$&    39.62$\%$ & 2784\\
& Men & 56.67$\%$ & 15.47$\%$ & 27.85$\%$ & 2592\\ 
	\cmidrule(lr){1-1} \cmidrule(lr){2-2} \cmidrule(lr){3-5}  \cmidrule(lr){6-6}
\multirow{2}{*}{PT} & Women &    39.66$\%$ &    15.63$\%$ &    44.72$\%$ & 2784\\
& Men & 39.08$\%$ & 21.53$\%$ & 39.39& 2592\\ 
	\cmidrule(lr){1-1} \cmidrule(lr){2-2} \cmidrule(lr){3-5}  \cmidrule(lr){6-6}
\multirow{2}{*}{All transfers} & Women &    49.16$\%$ &    12.37$\%$ &    38.47$\%$ & 9280\\
& Men & 54.79$\%$ & 16.19$\%$ & 29.02$\%$ & 8640 \\ 
 \bottomrule
\end{tabular}
\end{table}

\begin{table}[!htp]\caption{\textit{Equality tests of the acceptance rates by gender}} \label{Equality-gender}
    \vspace{0.2cm}
    \centering
    \footnotesize
\begin{tabular}{llccc}
\toprule
$\chi^{2}$ \textsc{Statistics} & & \textsc{DL} & \textsc{Value} & \textsc{Prob.} \\
	\cmidrule(lr){1-1} \cmidrule(lr){2-2} \cmidrule(lr){3-3}  \cmidrule(lr){4-5}
\multirow{2}{*}{Global}& Women & 3 & 213.09 & < 0.0001  \\
& Men & 3 & 438.33 & < 0.0001  \\ 
	\cmidrule(lr){1-1} \cmidrule(lr){2-2} \cmidrule(lr){3-3}  \cmidrule(lr){4-5}
\multirow{2}{*}{URL versus UR}& Women & 1 & 52.44 & < 0.0001  \\
& Men & 1 & 69.62 & < 0.0001 \\ 
	\cmidrule(lr){1-1} \cmidrule(lr){2-2} \cmidrule(lr){3-3}  \cmidrule(lr){4-5}
\multirow{2}{*}{URL versus UL}& Women & 1 & 7.51 & 0.006  \\
& Men & 1 & 28.72 & < 0.0001 \\ 
	\cmidrule(lr){1-1} \cmidrule(lr){2-2} \cmidrule(lr){3-3}  \cmidrule(lr){4-5}
\multirow{2}{*}{URL versus PT}& Women & 1 & 131.89 & < 0.0001  \\
& Men & 1 & 293.43 & < 0.0001  \\ 
	\cmidrule(lr){1-1} \cmidrule(lr){2-2} \cmidrule(lr){3-3}  \cmidrule(lr){4-5}
\multirow{2}{*}{UR versus UL}& Women & 1 & 41.09 & < 0.0001   \\
& Men & 1 & 19.26 & < 0.0001 \\ 
	\cmidrule(lr){1-1} \cmidrule(lr){2-2} \cmidrule(lr){3-3}  \cmidrule(lr){4-5}
\multirow{2}{*}{UR versus PT}& Women & 1 & 35.67 & < 0.0001  \\
& Men & 1 & 160.73 & < 0.0001 \\ 
	\cmidrule(lr){1-1} \cmidrule(lr){2-2} \cmidrule(lr){3-3}  \cmidrule(lr){4-5}
\multirow{2}{*}{UL versus PT}& Women & 1 & 152.28 & <  0.0001  \\
& Men & 1 & 288.14 & < 0.0001 \\ 
\bottomrule
\end{tabular}\\
    \vspace{0.2cm}
    \textit{Notes}. Null hypothesis $\rightarrow$ equality of the acceptance rates.
\end{table}

\begin{table}[!htp]\caption{\textit{Equality tests of the acceptance rates between gender}} \label{Equality-gender2}
    \vspace{0.2cm}
    \centering
    \footnotesize
\begin{tabular}{llccc}
\toprule
\multicolumn{2}{l}{$\chi^{2}$ \textsc{Statistics}}  & \textsc{DL} & \textsc{Value} & \textsc{Prob.} \\
	\cmidrule(lr){1-2} \cmidrule(lr){3-3} \cmidrule(lr){4-5}
URL & (Men/Women) & 1 & 26.08 & < 0.0001 \\
UL    & (Men/Women) & 1 & 23.33 & < 0.0001 \\
UR   & (Men/Women) & 1 & 44.35 & < 0.0001 \\
PT      & (Men/Women) & 1 & 0.18 & \textbf{0.667}\\
\bottomrule
\end{tabular}\\
    \vspace{0.2cm}
    \textit{Notes}. Null hypothesis $\rightarrow$ equality of the acceptance rates.
\end{table}

\begin{table}[!htp]\caption{\textit{Acceptation rates by level of education}} \label{Resulats-ecole}
    \vspace{0.2cm}
    \centering
    \footnotesize
\begin{tabular}{ll*{4}{c}}
\toprule
\textsc{Transfers} & & \textsc{Accepted} & \textsc{Rejected} & \textsc{Neutrality} & \textit{N} \\
	\cmidrule(lr){1-1} \cmidrule(lr){2-2} \cmidrule(lr){3-5}  \cmidrule(lr){6-6}
\multirow{4}{*}{URL} & Before high school &    45.83$\%$&    19.79$\%$&    34.38$\%$& 96\\
& High school & 60.98$\%$ & 11.38$\%$ & 27.64$\%$ & 492\\
& Short tertiary educ. & 70.15$\%$ & 8.93$\%$ & 20.92$\%$ & 392\\ 
& University degree & 71.18$\%$ & 5.05$\%$ & 23.77$\%$ & 812\\  
	\cmidrule(lr){1-1} \cmidrule(lr){2-2} \cmidrule(lr){3-5}  \cmidrule(lr){6-6}
\multirow{4}{*}{UL} & Before high school &    45.14$\%$ &    20.83$\%$ &    34.03$\%$ & 288\\
& High school & 53.73$\%$ & 15.51$\%$ & 30.76$\%$ & 1476\\
& Short tertiary educ. & 63.61$\%$ & 13.27$\%$ & 23.13$\%$ & 1176\\ 
& University degree & 62.27$\%$ & 8.25$\%$ & 29.47$\%$ & 2436\\  
	\cmidrule(lr){1-1} \cmidrule(lr){2-2} \cmidrule(lr){3-5}  \cmidrule(lr){6-6}
\multirow{4}{*}{UR} & Before high school &    35.42$\%$ &    21.18$\%$ &    43.40$\%$ & 288\\
& High school & 47.09$\%$ & 17.75$\%$ & 35.16$\%$ & 1476\\
& Short tertiary educ. & 56.04$\%$ & 14.46$\%$ & 29.51$\%$ & 1176\\ 
& University degree & 54.93$\%$ & 10.84$\%$ & 34.24$\%$ & 2436\\  
	\cmidrule(lr){1-1} \cmidrule(lr){2-2} \cmidrule(lr){3-5}  \cmidrule(lr){6-6}
\multirow{4}{*}{PT} & Before high school &    34.72$\%$ &    24.31$\%$ &    40.97$\%$ & 288\\
& High school & 36.11$\%$ & 23.17$\%$ & 40.72$\%$ & 1476\\
& Short tertiary educ. & 42.94$\%$ & 17.86$\%$ & 39.20$\%$ & 1176\\ 
& University degree & 40.19$\%$ & 15.23$\%$ & 44.58$\%$ & 2436\\  
	\cmidrule(lr){1-1} \cmidrule(lr){2-2} \cmidrule(lr){3-5}  \cmidrule(lr){6-6}
\multirow{4}{*}{All transfers} & Before high school &    39.17$\%$ &    21.88$\%$ &    38.96$\%$ & 960\\
& High school & 47.17$\%$ & 18.07$\%$ & 34.76$\%$ & 4920\\
& Short tertiary educ. & 55.79$\%$ & 14.57$\%$ & 29.64$\%$ & 3920\\ 
& University degree & 54.33$\%$ & 10.80$\%$ & 34.86$\%$ & 8120\\  
 \bottomrule
\end{tabular}
\end{table}

\begin{table}[!htp]\caption{\textit{Equality tests of the acceptance rates by level of education}} \label{Equality-schooling}
    \vspace{0.2cm}
    \centering
    \footnotesize
\begin{tabular}{llccc}
\toprule
$\chi^{2}$ \textsc{Statistics} & & \textsc{DL} & \textsc{Value} & \textsc{Prob.} \\
	\cmidrule(lr){1-1} \cmidrule(lr){2-2} \cmidrule(lr){3-3}  \cmidrule(lr){4-5}
\multirow{4}{*}{Global}& Before high school & 3 & 10.19 & 0.017 \\
& High school & 3 & 135.53 & < 0.0001   \\ 
& Short tertiary educ. & 3 & 140.64 & < 0.0001   \\ 
& University degree & 3 & 351.58 & < 0.0001   \\ 
	\cmidrule(lr){1-1} \cmidrule(lr){2-2} \cmidrule(lr){3-3}  \cmidrule(lr){4-5}
\multirow{4}{*}{URL versus UR}& Before high school & 1 & 3.32 & \textbf{0.069}  \\
& High school & 1 & 28.48 & < 0.0001   \\ 
& Short tertiary educ. & 1 & 24.32 & < 0.0001   \\ 
& University degree & 1 & 66.53 & < 0.0001   \\  
	\cmidrule(lr){1-1} \cmidrule(lr){2-2} \cmidrule(lr){3-3}  \cmidrule(lr){4-5}
\multirow{4}{*}{URL versus UL}& Before high school & 1 & 0.01 & \textbf{0.906}  \\
& High school & 1 & 7.85 & 0.005   \\ 
& Short tertiary educ. & 1 & 5.56 &  0.018   \\ 
& University degree & 1 & 21.11 & < 0.0001   \\  
	\cmidrule(lr){1-1} \cmidrule(lr){2-2} \cmidrule(lr){3-3}  \cmidrule(lr){4-5}
\multirow{4}{*}{URL versus PT}& Before high school & 1 & 3.79 & \textbf{0.051}  \\
& High school & 1 & 93.45 & < 0.0001   \\ 
& Short tertiary educ. & 1 & 87.08 & < 0.0001   \\ 
& University degree & 1 & 234.40 & < 0.0001   \\  
	\cmidrule(lr){1-1} \cmidrule(lr){2-2} \cmidrule(lr){3-3}  \cmidrule(lr){4-5}
\multirow{4}{*}{UR versus UL}& Before high school & 1 & 5.66 & 0.017  \\
& High school & 1 & 13.01 & < 0.0001   \\ 
& Short tertiary educ. & 1 & 14.01 & < 0.0001   \\ 
& University degree & 1 & 27.11 & < 0.0001   \\  
	\cmidrule(lr){1-1} \cmidrule(lr){2-2} \cmidrule(lr){3-3}  \cmidrule(lr){4-5}
\multirow{4}{*}{UR versus PT}& Before high school & 1 & 0.03 & \textbf{0.861}  \\
& High school & 1 & 36.60 & < 0.0001   \\ 
& Short tertiary educ. & 1 & 40.34 & < 0.0001   \\ 
& University degree & 1 & 106.07 & < 0.0001   \\  
	\cmidrule(lr){1-1} \cmidrule(lr){2-2} \cmidrule(lr){3-3}  \cmidrule(lr){4-5}
\multirow{4}{*}{UL versus PT}& Before high school & 1 & 6.51 & 0.011  \\
& High school & 1 & 92.55 & < 0.0001   \\ 
& Short tertiary educ. & 1 & 100.86 & < 0.0001   \\ 
& University degree & 1 & 237.78 & < 0.0001   \\   
\bottomrule
\end{tabular}\\
    \vspace{0.2cm}
    \textit{Notes}. Null hypothesis $\rightarrow$ equality of the acceptance rates.
\end{table}

\begin{table}[!htp]\caption{\textit{Equality tests of the acceptance rates between level of education}} \label{Equality-schooling2}
    \vspace{0.2cm}
    \centering
    \footnotesize
\begin{tabular}{llccc}
\toprule
\multicolumn{2}{l}{$\chi^{2}$ \textsc{Statistics}}  & \textsc{DL} & \textsc{Value} & \textsc{Prob.} \\
	\cmidrule(lr){1-2} \cmidrule(lr){3-3} \cmidrule(lr){4-5}
URL & (Before high school/High school/\dots/Univ. degree) & 3 & 35.57 & < 0.0001 \\
UL    & (Before high school/High school/\dots/Univ. degree) & 3 & 60.89 & < 0.0001 \\
UR   & (Before high school/High school/\dots/Univ. degree) & 3 & 62.04 & < 0.0001 \\
PT    & (Before high school/High school/\dots/Univ. degree) & 3 & 16.14 & 0.001\\
\bottomrule
\end{tabular}\\
    \vspace{0.2cm}
    \textit{Notes}. Null hypothesis $\rightarrow$ equality of the acceptance rates.
\end{table}

\begin{table}[!htp]\caption{\textit{Acceptation rates by political views}} \label{Resulats-politique}
    \vspace{0.2cm}
    \centering
    \footnotesize
\begin{tabular}{ll*{4}{c}}
\toprule
\textsc{Transfers} & & \textsc{Accepted} & \textsc{Rejected} & \textsc{Neutrality} & \textit{N} \\
	\cmidrule(lr){1-1} \cmidrule(lr){2-2} \cmidrule(lr){3-5}  \cmidrule(lr){6-6}
\multirow{6}{*}{URL} & Wish not to answer &    59.34$\%$ &    12.37$\%$ &    28.28$\%$ & 396\\
& Far Left & 73.33$\%$ & 3.33$\%$ & 23.33$\%$ & 60\\
& Left & 70.25$\%$ & 5.58$\%$ & 24.17$\%$ & 484\\ 
& Centre & 70.83$\%$ & 8.33$\%$ & 20.83$\%$ & 480\\
& Right & 64.38$\%$ & 7.53$\%$ & 28.08$\%$ & 292\\
& Far Right & 62.50$\%$ & 13.75$\%$ & 23.75$\%$ & 80\\  
	\cmidrule(lr){1-1} \cmidrule(lr){2-2} \cmidrule(lr){3-5}  \cmidrule(lr){6-6}
\multirow{6}{*}{UL} & Wish not to answer &    54.38$\%$ &    14.90$\%$ &    30.72$\%$ & 1188\\
& Far Left & 65.00$\%$ & 4.44$\%$ & 30.56$\%$ & 180\\
& Left & 60.95$\%$ & 9.64$\%$ & 29.41$\%$ & 1452\\ 
& Centre & 60.90$\%$ & 13.19$\%$ & 25.90$\%$ & 1440\\
& Right & 60.39$\%$ & 10.73$\%$ & 28.88$\%$ & 876\\
& Far Right & 55.83$\%$ & 15.42$\%$ & 28.75$\%$ & 240\\   
	\cmidrule(lr){1-1} \cmidrule(lr){2-2} \cmidrule(lr){3-5}  \cmidrule(lr){6-6}
\multirow{6}{*}{UR} & Wish not to answer &    46.72$\%$ &    15.82$\%$ &    37.46$\%$ & 1188\\
& Far Left & 47.78$\%$ & 5.56$\%$ & 46.67$\%$ & 180\\
& Left & 54.89$\%$ & 12.47$\%$ & 32.64$\%$ & 1452\\ 
& Centre & 53.96$\%$ & 15.00$\%$ & 31.04$\%$ & 1440\\
& Right & 51.14$\%$ & 13.01$\%$ & 35.84$\%$ & 876\\
& Far Right & 54.58$\%$ & 20.00$\%$ & 25.42$\%$ & 240\\   
	\cmidrule(lr){1-1} \cmidrule(lr){2-2} \cmidrule(lr){3-5}  \cmidrule(lr){6-6}
\multirow{6}{*}{PT} & Wish not to answer &    33.75$\%$ &    18.86$\%$ &    47.39$\%$ & 1188\\
& Far Left & 48.33$\%$ & 7.22$\%$ & 44.44$\%$ & 180\\
& Left & 38.36$\%$ & 17.29$\%$ & 44.35$\%$ & 1452\\ 
& Centre & 40.90$\%$ & 20.83$\%$ & 38.26$\%$ & 1440\\
& Right & 42.81$\%$ & 15.98$\%$ & 41.21$\%$ & 876\\
& Far Right & 45.00$\%$ & 27.08$\%$ & 27.92$\%$ & 240\\   
	\cmidrule(lr){1-1} \cmidrule(lr){2-2} \cmidrule(lr){3-5}  \cmidrule(lr){6-6}
\multirow{6}{*}{All transfers} & Wish not to answer &    46.39$\%$ &    16.11$\%$ &    37.50$\%$ & 3960\\
& Far Left & 55.67$\%$ & 5.50$\%$ & 38.83$\%$ & 600\\
& Left & 53.29$\%$ & 12.38$\%$ & 34.34$\%$ & 4840\\ 
& Centre & 53.81$\%$ & 15.54$\%$ & 30.65$\%$ & 4800\\
& Right & 52.74$\%$ & 12.67$\%$ & 34.59$\%$ & 2920\\
& Far Right & 52.88$\%$ & 20.13$\%$ & 27.00$\%$ & 800\\   
 \bottomrule
\end{tabular}
\end{table}

\begin{table}[!htp]\caption{\textit{Equality tests of the acceptance rates by political views}} \label{Equality-politique}
    \vspace{0.2cm}
    \centering
    \footnotesize
\begin{tabular}{llccc}
\toprule
$\chi^{2}$ \textsc{Statistics} & & \textsc{DL} & \textsc{Value} & \textsc{Prob.} \\
	\cmidrule(lr){1-1} \cmidrule(lr){2-2} \cmidrule(lr){3-3}  \cmidrule(lr){4-5}
\multirow{6}{*}{Global}& Wish not to answer & 3 & 133.51 & < 0.0001 \\
& Far left & 3 & 22.40 & < 0.0001   \\ 
& Left & 3 & 221.65 & < 0.0001   \\ 
& Centre & 3 & 181.65 & < 0.0001   \\
& Right & 3 & 72.01 & < 0.0001   \\ 
& Far Right & 3 & 10.07 & < 0.0001   \\  
	\cmidrule(lr){1-1} \cmidrule(lr){2-2} \cmidrule(lr){3-3}  \cmidrule(lr){4-5}
\multirow{6}{*}{URL versus UR}& Wish not to answer & 1 & 18.94 & < 0.0001 \\
& Far left & 1 & 11.84 & 0.001   \\ 
& Left & 1 & 35.33 & < 0.0001   \\ 
& Centre & 1 & 42.13 & < 0.0001   \\
& Right & 1 & 15.48 & < 0.0001   \\ 
& Far Right & 1 & 1.53 & \textbf{0.216}   \\    
	\cmidrule(lr){1-1} \cmidrule(lr){2-2} \cmidrule(lr){3-3}  \cmidrule(lr){4-5}
\multirow{6}{*}{URL versus UL}& Wish not to answer & 1 & 2.97  & \textbf{0.085} \\
& Far left & 1 & 1.42 & \textbf{0.234}   \\ 
& Left & 1 & 13.50 & < 0.0001   \\ 
& Centre & 1 & 15.30 & < 0.0001   \\
& Right & 1 & 1.47 &  \textbf{0.225}   \\ 
& Far Right & 1 & 1.09 & \textbf{0.296}   \\      
	\cmidrule(lr){1-1} \cmidrule(lr){2-2} \cmidrule(lr){3-3}  \cmidrule(lr){4-5}
\multirow{6}{*}{URL versus PT}& Wish not to answer & 1 & 80.93 & < 0.0001 \\
& Far left & 1 & 11.35 & 0.001   \\ 
& Left & 1 & 148.44 & < 0.0001   \\ 
& Centre & 1 & 129.14 & < 0.0001   \\
& Right & 1 & 40.83 & < 0.0001   \\ 
& Far Right & 1 & 7.35 & 0.007   \\     
	\cmidrule(lr){1-1} \cmidrule(lr){2-2} \cmidrule(lr){3-3}  \cmidrule(lr){4-5}
\multirow{6}{*}{UR versus UL}& Wish not to answer & 1 & 13.94 & < 0.0001 \\
& Far left & 1 & 10.86 & 0.001   \\ 
& Left & 1 & 10.94 & 0.001   \\ 
& Centre & 1 & 14.20 & < 0.0001   \\
& Right & 1 & 15.18 & < 0.0001   \\ 
& Far Right & 1 & 0.08 & \textbf{0.783}   \\      
	\cmidrule(lr){1-1} \cmidrule(lr){2-2} \cmidrule(lr){3-3}  \cmidrule(lr){4-5}
\multirow{6}{*}{UR versus PT}& Wish not to answer & 1 & 41.51 & < 0.0001 \\
& Far left & 1 & 0.01 &  \textbf{0.916}   \\ 
& Left & 1 & 79.70 & < 0.0001   \\ 
& Centre & 1 & 49.22 & < 0.0001   \\
& Right & 1 & 12.21 & < 0.0001   \\ 
& Far Right & 1 & 4.41 & 0.036   \\     
	\cmidrule(lr){1-1} \cmidrule(lr){2-2} \cmidrule(lr){3-3}  \cmidrule(lr){4-5}
\multirow{6}{*}{UL versus PT}& Wish not to answer & 1 & 102.50 & < 0.0001 \\
& Far left & 1 & 10.18 & 0.001   \\ 
& Left & 1 & 148.19 & < 0.0001   \\ 
& Centre & 1 & 115.24 & < 0.0001   \\
& Right & 1 & 54.20 & < 0.0001   \\ 
& Far Right & 1 & 5.63 & 0.018   \\       
\bottomrule
\end{tabular}\\
    \vspace{0.2cm}
    \textit{Notes}. Null hypothesis $\rightarrow$ equality of the acceptance rates.
\end{table}

\begin{table}[!htp]\caption{\textit{Equality tests of the acceptance rates between political views}} \label{Equality-politique2}
    \vspace{0.2cm}
    \centering
    \footnotesize
\begin{tabular}{llccc}
\toprule
\multicolumn{2}{l}{$\chi^{2}$ \textsc{Statistics}}  & \textsc{DL} & \textsc{Value} & \textsc{Prob.} \\
	\cmidrule(lr){1-2} \cmidrule(lr){3-3} \cmidrule(lr){4-5}
URL & (No answer/Far Left/\dots/F. right) & 5 &  18.63 & 0.002 \\
UL    & (No answer/Far Left/\dots/F. right) & 5 & 19.15 & 0.002 \\
UR   & (No answer/Far Left/\dots/F. right) & 5 & 22.54 & < 0.0001 \\
PT    & (No answer/Far Left/\dots/F. right) & 5 & 31.31 & < 0.0001\\
\bottomrule
\end{tabular}\\
    \vspace{0.2cm}
    \textit{Notes}. Null hypothesis $\rightarrow$ equality of the acceptance rates.
\end{table}

\begin{table}[!htp]\caption{\textit{Acceptation rates by employment status}} \label{Resulats-profession}
    \vspace{0.2cm}
    \centering
    \footnotesize
\begin{tabular}{ll*{4}{c}}
\toprule
\textsc{Transfers} & & \textsc{Accepted} & \textsc{Rejected} & \textsc{Neutrality} & \textit{N} \\
	\cmidrule(lr){1-1} \cmidrule(lr){2-2} \cmidrule(lr){3-5}  \cmidrule(lr){6-6}
\multirow{7}{*}{URL} & Full-time &    66.76$\%$ &    6.39$\%$ &    26.85$\%$ & 704\\
& Part-time & 76.39$\%$ & 5.56$\%$ & 18.06$\%$ & 72\\
& Self-employed & 50.93$\%$ & 12.96 & 36.11$\%$ & 108\\ 
& Seeking employment & 70.00$\%$ & 14.00$\%$ & 16.00$\%$ & 100\\
& Student & 55.15$\%$ & 12.50$\%$ & 32.35$\%$ & 136\\
& Unempl. not seeking & 58.93$\%$ & 8.93$\%$ & 32.14$\%$ & 112\\
& Retired & 72.50$\%$ & 8.39$\%$ & 19.11$\%$ & 560\\   
	\cmidrule(lr){1-1} \cmidrule(lr){2-2} \cmidrule(lr){3-5}  \cmidrule(lr){6-6}
\multirow{7}{*}{UL} & Full-time &    57.58$\%$ &    11.08$\%$ &    31.34$\%$ & 2112\\
& Part-time & 69.91$\%$ & 7.41$\%$ & 22.69$\%$ & 216\\
& Self-employed & 47.53$\%$ & 16.05$\%$ & 36.42$\%$ & 324\\ 
& Seeking employment & 62.00$\%$ & 17.67$\%$ & 20.33$\%$ & 300\\
& Student & 49.02$\%$ & 15.69$\%$ & 35.29$\%$ & 408\\
& Unempl. not seeking & 55.06$\%$ & 10.42$\%$ & 34.52$\%$ & 336\\
& Retired & 65.24$\%$ & 11.43$\%$ & 23.33$\%$ & 1680\\    
	\cmidrule(lr){1-1} \cmidrule(lr){2-2} \cmidrule(lr){3-5}  \cmidrule(lr){6-6}
\multirow{7}{*}{UR} & Full-time &    50.05$\%$ &    14.54$\%$ &    35.42$\%$ & 2112\\
& Part-time & 61.11$\%$ & 14.35$\%$ & 24.54$\%$ & 216\\
& Self-employed & 33.64$\%$ & 16.67$\%$ & 49.69$\%$ & 324\\ 
& Seeking employment & 53.00$\%$ & 17.67$\%$ & 29.33$\%$ & 300\\
& Student & 43.14$\%$ & 16.67$\%$ & 40.20$\%$ & 408\\
& Unempl. not seeking & 49.40$\%$ & 13.39$\%$ & 37.20$\%$ & 336\\
& Retired & 59.23$\%$ & 11.85$\%$ & 28.93$\%$ & 1680\\    
	\cmidrule(lr){1-1} \cmidrule(lr){2-2} \cmidrule(lr){3-5}  \cmidrule(lr){6-6}
\multirow{7}{*}{PT} & Full-time &    37.64$\%$ &    17.19$\%$ &    45.17$\%$ & 2112\\
& Part-time & 35.65$\%$ & 25.93$\%$ & 38.43$\%$ & 216\\
& Self-employed & 28.70$\%$ & 20.37$\%$ & 50.93$\%$ & 324\\ 
& Seeking employment & 42.00$\%$ & 22.67$\%$ & 35.33$\%$ & 300\\
& Student & 31.37$\%$ & 21.08$\%$ & 47.55$\%$ & 408\\
& Unempl. not seeking & 46.13$\%$ & 15.77$\%$ & 38.10$\%$ & 336\\
& Retired & 44.23$\%$ & 17.92$\%$ & 37.86$\%$ & 1680\\    
	\cmidrule(lr){1-1} \cmidrule(lr){2-2} \cmidrule(lr){3-5}  \cmidrule(lr){6-6}
\multirow{7}{*}{All transfers} & Full-time &    50.26$\%$ &    13.48$\%$ &    36.26$\%$ & 7040\\
& Part-time & 57.64$\%$ & 14.86$\%$ & 27.50$\%$ & 720\\
& Self-employed & 38.06$\%$ & 17.22$\%$ & 44.72$\%$ & 1080\\ 
& Seeking employment & 54.10$\%$ & 18.80$\%$ & 27.10$\%$ & 1000\\
& Student & 42.57$\%$ & 17.28$\%$ & 40.15$\%$ & 1360\\
& Unempl. not seeking & 51.07$\%$ & 12.77$\%$ & 36.16$\%$ & 1120\\
& Retired & 57.86$\%$ & 13.20$\%$ & 28.95$\%$ & 5600\\    
 \bottomrule
\end{tabular}
\end{table}

\begin{table}[!htp]\caption{\textit{Equality tests of the acceptance rates by employment status}} \label{Equality-profession}
    \vspace{0.2cm}
    \centering
    \scriptsize
\begin{tabular}{llccc}
\toprule
$\chi^{2}$ \textsc{Statistics} & & \textsc{DL} & \textsc{Value} & \textsc{Prob.} \\
	\cmidrule(lr){1-1} \cmidrule(lr){2-2} \cmidrule(lr){3-3}  \cmidrule(lr){4-5}
\multirow{7}{*}{Global}& Full-time & 3 & 256.44 & < 0.0001 \\
& Part-time & 3 & 67.53 & < 0.0001   \\ 
& Self-employed & 3 & 34.63 & < 0.0001   \\ 
& Seeking employment & 3 & 35.56 & < 0.0001   \\
& Student & 3 & 36.72 & < 0.0001   \\ 
& Unempl. not seeking & 3 &  8.56  &  0.036   \\ 
& Retired & 3 & 216.09 & < 0.0001   \\  
	\cmidrule(lr){1-1} \cmidrule(lr){2-2} \cmidrule(lr){3-3}  \cmidrule(lr){4-5}
\multirow{7}{*}{URL versus UR}& Full-time & 1 & 59.43 & < 0.0001 \\
& Part-time & 1 & 5.54 & 0.019   \\ 
& Self-employed & 1 & 10.27 & 0.001   \\ 
& Seeking employment & 1 & 8.86 & 0.003   \\
& Student & 1 & 5.92 &  0.015   \\ 
& Unempl. not seeking & 1 & 3.05 &  \textbf{0.081}   \\ 
& Retired & 1 & 31.59 & < 0.0001   \\      
	\cmidrule(lr){1-1} \cmidrule(lr){2-2} \cmidrule(lr){3-3}  \cmidrule(lr){4-5}
\multirow{7}{*}{URL versus UL}& Full-time & 1 & 18.54 & < 0.0001 \\
& Part-time & 1 & 1.11 &  \textbf{0.291}   \\ 
& Self-employed & 1 & 0.37 & \textbf{0.541}   \\ 
& Seeking employment & 1 & 2.08 & \textbf{0.149}   \\
& Student & 1 & 1.53 & \textbf{0.216}   \\ 
& Unempl. not seeking & 1 & 0.51 & \textbf{0.475}   \\ 
& Retired & 1 & 10.03 & 0.002   \\     
	\cmidrule(lr){1-1} \cmidrule(lr){2-2} \cmidrule(lr){3-3}  \cmidrule(lr){4-5}
\multirow{7}{*}{URL versus PT}& Full-time & 1 &  180.95 & < 0.0001 \\
& Part-time & 1 & 36.10 & < 0.0001   \\ 
& Self-employed & 1 & 17.76 & < 0.0001   \\ 
& Seeking employment & 1 & 23.53 & < 0.0001   \\
& Student & 1 & 24.65 & < 0.0001   \\ 
& Unempl. not seeking & 1 & 5.50 &  0.019   \\ 
& Retired & 1 & 134.39 & < 0.0001   \\        
	\cmidrule(lr){1-1} \cmidrule(lr){2-2} \cmidrule(lr){3-3}  \cmidrule(lr){4-5}
\multirow{7}{*}{UR versus UL}& Full-time & 1 & 24.08 & < 0.0001 \\
& Part-time & 1 &  3.70 &  \textbf{0.054}   \\ 
& Self-employed & 1 & 12.96 & < 0.0001   \\ 
& Seeking employment & 1 & 4.97 &  0.026   \\
& Student & 1 & 2.84 & \textbf{0.092}   \\ 
& Unempl. not seeking & 1 & 2.15 & \textbf{0.142}   \\ 
& Retired & 1 & 12.92 & < 0.0001   \\         
	\cmidrule(lr){1-1} \cmidrule(lr){2-2} \cmidrule(lr){3-3}  \cmidrule(lr){4-5}
\multirow{7}{*}{UR versus PT}& Full-time & 1 & 66.00 & < 0.0001 \\
& Part-time & 1 & 28.04 & < 0.0001   \\ 
& Self-employed & 1 & 1.84 & \textbf{0.175}   \\ 
& Seeking employment & 1 & 7.28 & 0.007   \\
& Student & 1 &  12.08 & 0.001   \\ 
& Unempl. not seeking & 1 &  0.72 &  \textbf{0.396}   \\ 
& Retired & 1 & 75.69 & < 0.0001   \\        
	\cmidrule(lr){1-1} \cmidrule(lr){2-2} \cmidrule(lr){3-3}  \cmidrule(lr){4-5}
\multirow{7}{*}{UL versus PT}& Full-time & 1 & 168.22 & < 0.0001 \\
& Part-time & 1 & 50.86 & < 0.0001   \\ 
& Self-employed & 1 & 24.34 & < 0.0001   \\ 
& Seeking employment & 1 & 24.04 & < 0.0001   \\
& Student & 1 & 26.43 & < 0.0001   \\ 
& Unempl. not seeking & 1 & 5.36 & 0.021   \\ 
& Retired & 1 & 149.68 & < 0.0001   \\        
\bottomrule
\end{tabular}\\
    \vspace{0.2cm}
    \textit{Notes}. Null hypothesis $\rightarrow$ equality of the acceptance rates.
\end{table}

\begin{table}[!htp]\caption{\textit{Equality tests of the acceptance rates between employment status}} \label{Equality-profession2}
    \vspace{0.2cm}
    \centering
    \footnotesize
\begin{tabular}{llccc}
\toprule
\multicolumn{2}{l}{$\chi^{2}$ \textsc{Statistics}}  & \textsc{DL} & \textsc{Value} & \textsc{Prob.} \\
	\cmidrule(lr){1-2} \cmidrule(lr){3-3} \cmidrule(lr){4-5}
URL & (Full-time/Part-time/\dots/Retired) & 6 & 35.38    & < 0.0001 \\
UL    & (Full-time/Part-time/\dots/Retired) & 6 & 77.09   & < 0.0001 \\
UR   & (Full-time/Part-time/\dots/Retired) & 6 & 103.16 & < 0.0001 \\
PT    & (Full-time/Part-time/\dots/Retired) & 6 & 54.17   & < 0.0001\\
\bottomrule
\end{tabular}\\
    \vspace{0.2cm}
    \textit{Notes}. Null hypothesis $\rightarrow$ equality of the acceptance rates.
\end{table}

\newpage
\subsection{Econometric estimates}
\label{app-kernel}

\begin{figure}[!htp] \caption{\label{Kernel-U} \textit{Kernel density of $\epsilon$ for the utility function $u_\epsilon$}} 
\vspace{0.2cm}
\small
\begin{tabular}{cc}
\includegraphics[width=8.5cm]{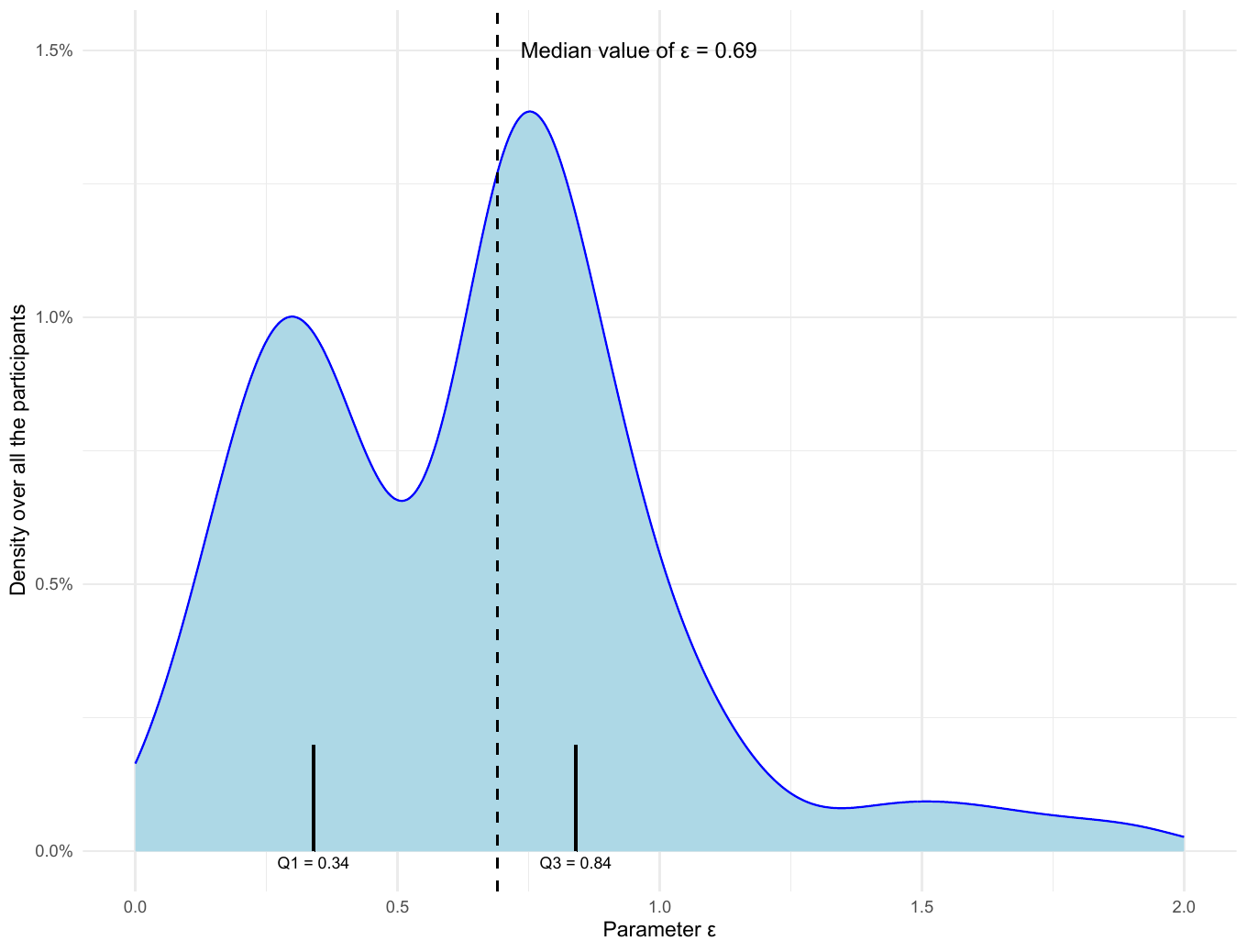} & %
\includegraphics[width=8.5cm]{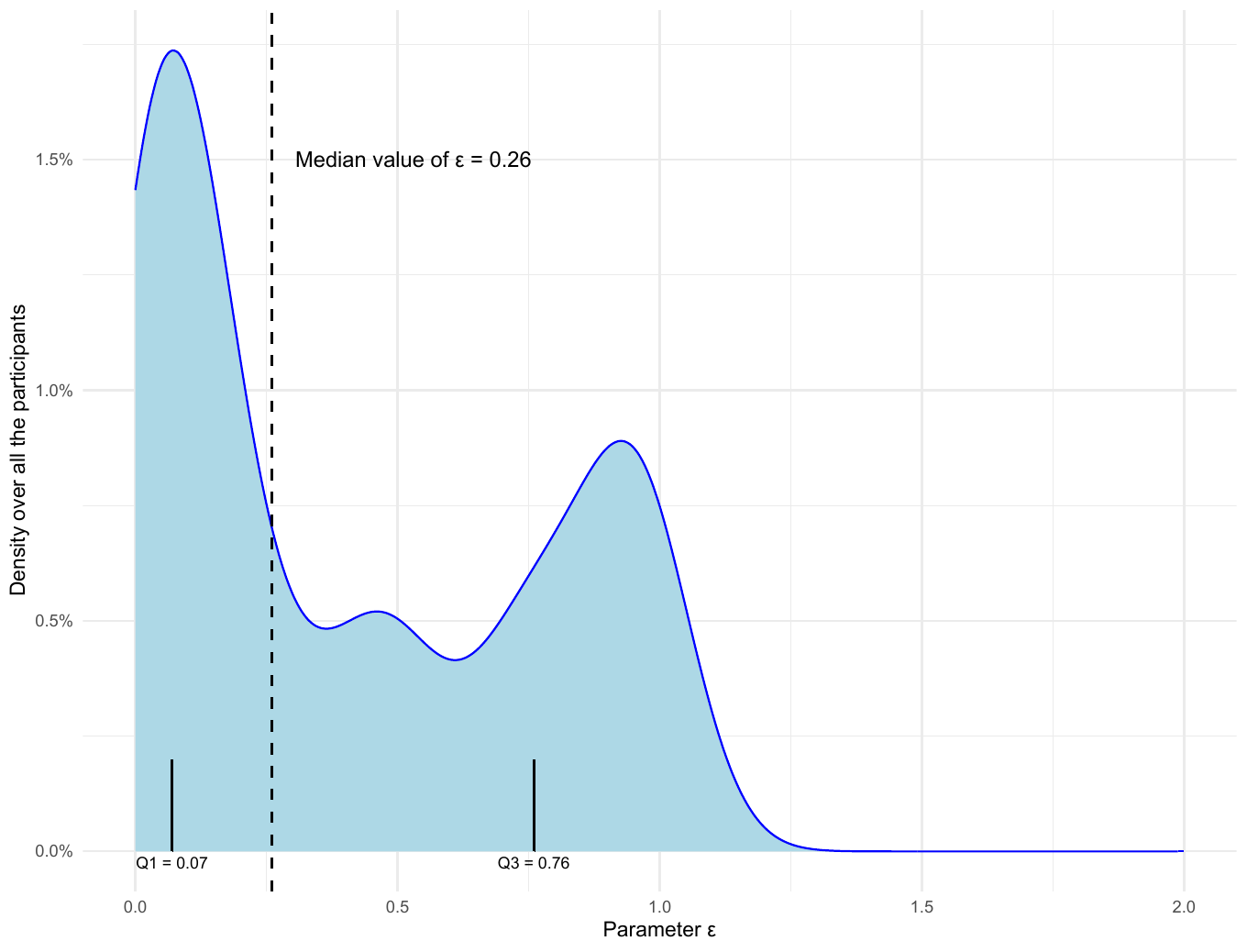} \\ 
\textit{SANN Algorithm} & \textit{BFGS Algorithm} 
\end{tabular}
\end{figure} 

\begin{figure}[H] \caption{\label{Kernel-F} \textit{Kernel density estimation of $\eta$ for the weighting function $f_\eta$}} 
\vspace{0.2cm}
\small
\begin{tabular}{cc}
\includegraphics[width=8.5cm]{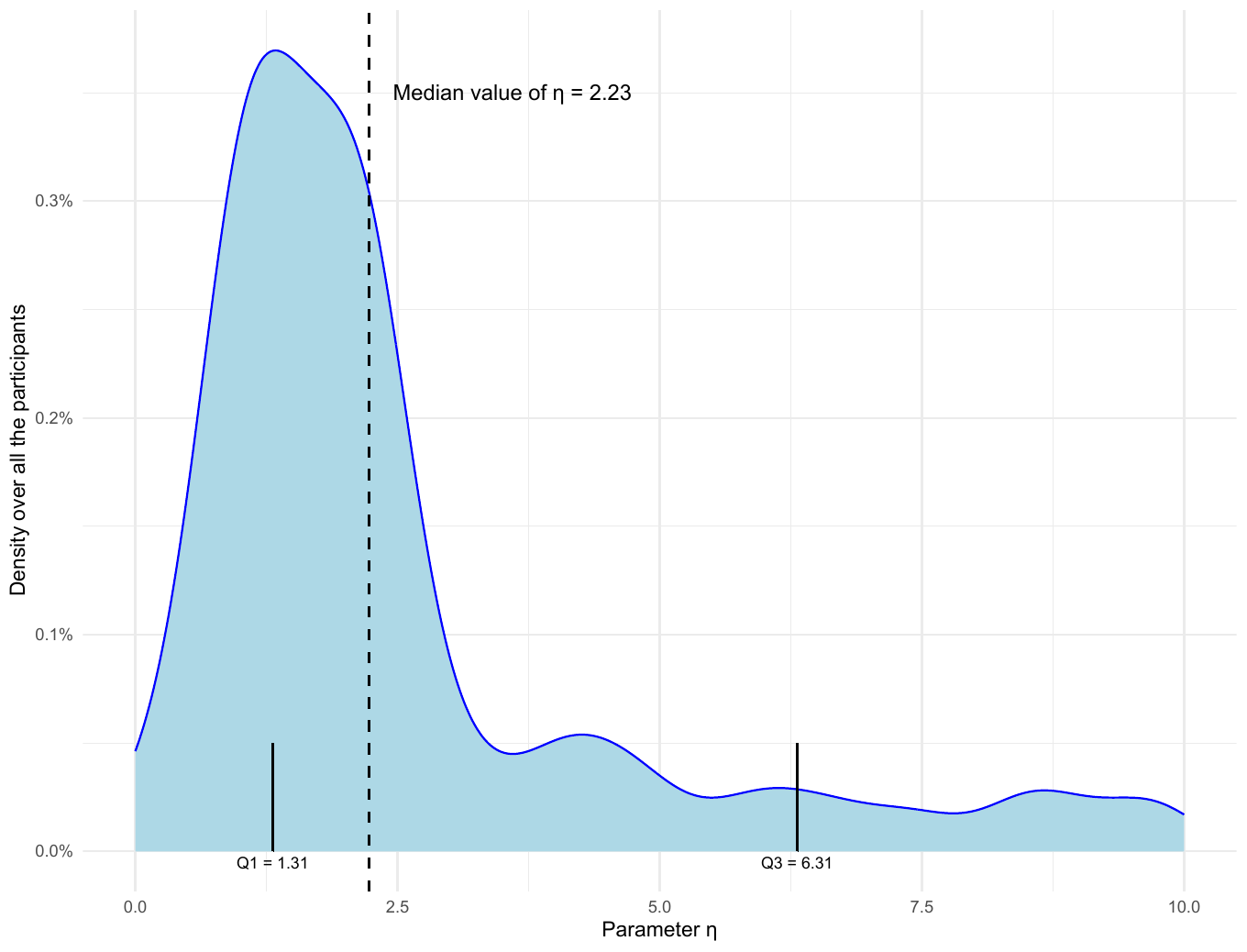} & %
\includegraphics[width=8.5cm]{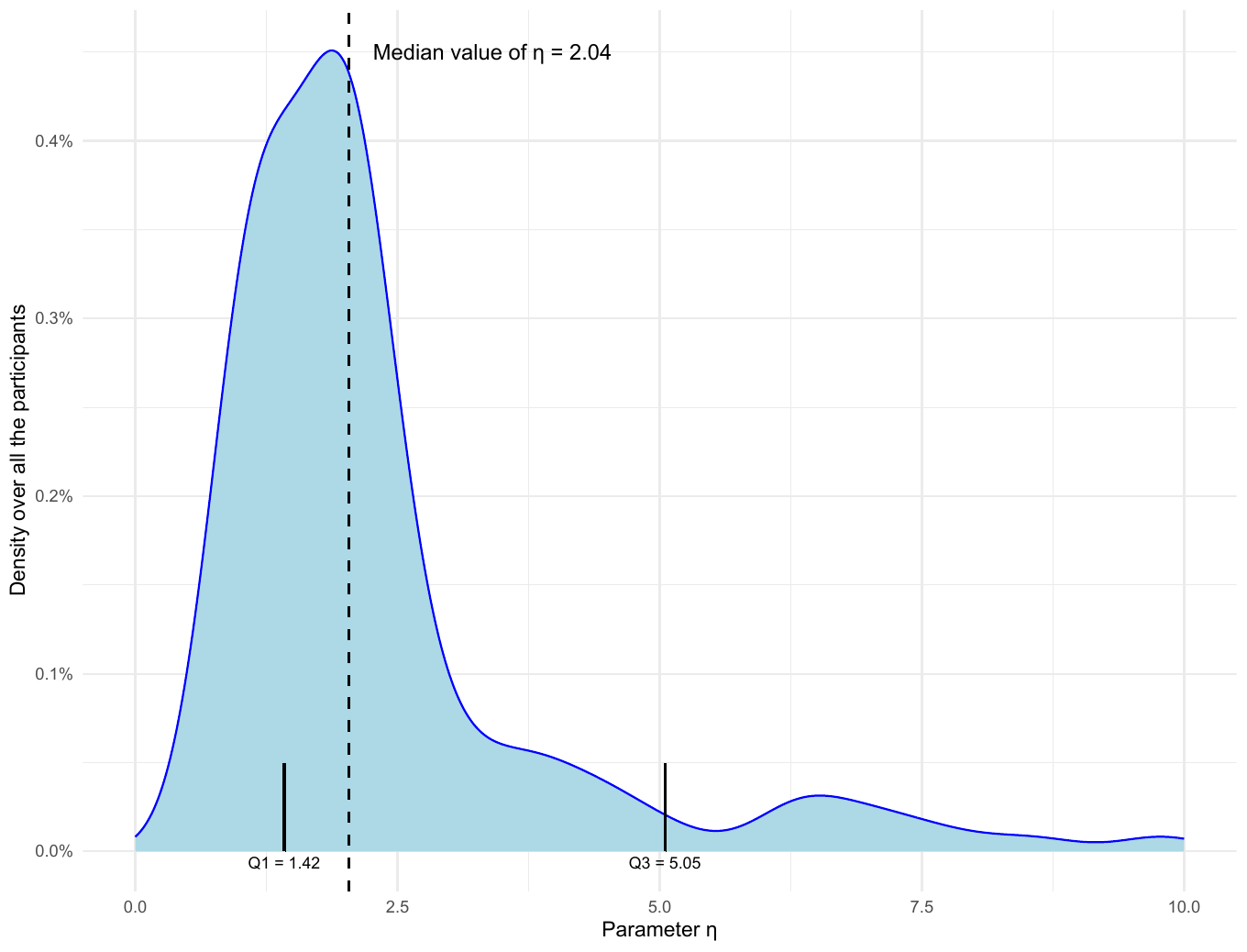} \\ 
\textit{SANN Algorithm} & \textit{BFGS Algorithm} 
\end{tabular}
\end{figure} 

\end{document}